\documentclass[11pt]{elsarticle}
\usepackage{epsfig,amsfonts,amssymb,amsmath}
\usepackage{color}
\usepackage{array}
\usepackage{lineno}
\usepackage{multirow}

\newtheorem{theorem}{Theorem}[section]

\newtheorem{lemma}[theorem]{Lemma}

\makeatletter
\def\ps@pprintTitle{%
  \let\@oddhead\@empty
  \let\@evenhead\@empty
  \let\@oddfoot\@empty
  \let\@evenfoot\@oddfoot
}
\makeatother

\begin{document}

\begin{frontmatter}

\title{Oscillations in epidemic models with spread of awareness}

\author[address1]{Winfried Just}
\ead{mathjust@gmail.com}

\author[address2]{Joan Salda\~{n}a\corref{cor1}}
\ead{joan.saldana@udg.edu}
\cortext[cor1]{Corresponding author}

\author[address1]{Ying Xin}
\ead{yx123812@ohio.edu}

\address[address1]{Department of Mathematics, Ohio University, Athens, OH 45701, U.S.A.}

\address[address2]{Departament d'Inform\`{a}tica, Matem\`{a}tica Aplicada i Estad\'istica
\\
Universitat de Girona, Catalonia, Spain}

\begin{abstract}
We study ODE models of epidemic spreading with a preventive behavioral response that is triggered by awareness of the infection. Previous studies of such models have mostly focused on  the impact of the response on the initial growth of an outbreak and the existence and location of endemic equilibria. Here we study the question whether this type of response is sufficient to prevent future flare-ups from low endemic levels if awareness is assumed to decay over time.  In the ODE context, such flare-ups would translate into sustained oscillations with significant amplitudes.

Our results show that such oscillations are ruled out in Susceptible-Aware-Infectious-Susceptible models with a single compartment of aware hosts, but can occur if we consider two distinct compartments of aware hosts who differ in their willingness to alert other susceptible hosts.
\end{abstract}


\end{frontmatter}

\section{Introduction}\label{s1}

The impact of behavioral responses on the progress of an infectious disease has received growing attention in epidemic modeling in recent years \cite{BLS,EPCH,Ferguson,FSJ}. Responses arising from risk perception of the disease are, for instance, avoidance behavior implying breaking off connections with infectious acquaintances (social distancing), and preventive behaviors, like handwashing or wearing face masks \cite{Kuo,Lau}. In epidemic network models, which take into account the contact network in a population, social avoidance has been modeled by means of several mechanisms of dynamic rewiring \cite{Gross06,JRS,Llensa,RZ,SRM}. Other approaches cast preventive actions into classic epidemic models by dividing each of the susceptible/infectious/removed classes between responsive and non-responsive individuals \cite{Funk10,Kiss10}, or by explicitly considering a new class of individuals who are both susceptible and aware, with a lowered susceptibility to infection relative to unaware hosts. This approach leads to the class of \emph{SAIS models} \cite{Juher14,Sahneh12a,Sahneh14,Sahneh11}.

Following \cite{Funk09,Funk10}, among others,  here we will consider \emph{awareness} as a state of knowledge about the prevalence of the disease that will both induce a behavioral response in the given host and that this host is willing to transmit to other hosts. To distinguish this more stringent notion of awareness from the one made in the SAIS models of the previous paragraph, we call the latter type of models \textit{reactive} SAIS models. These models incorporate direct transmission of information about the disease, so that the spread of awareness is somewhat similar but not entirely analogous to the spread of an infectious disease. See Section \ref{SAIS-Non} for details.

We are interested in two aspects of the question under what circumstances a behavioral response that is induced by awareness can be an effective control measure: whether such models would predict an elevated epidemic threshold and whether the response would prevent future flare-ups from low endemic levels. Some recent empirical findings appear to show such flare-ups when awareness or at the least adoption of the corresponding behavioral response diminishes. Several public health agencies have reported an increase in some sexual transmitted infections (STIs), particularly among MSM (men who have sex with men)  in high-income countries, since the mid-1990s. For instance, the Public Health Agency of Canada reports that syphilis infections are increasing in Canada and, between 2003 and 2012, rates increased by 100\% \cite{Totten}. Similarly, during the same period of time, the overall rates for gonorrhea and chlamydia in Canada increased by 39\% and 58\% (75\% among men), respectively (see \cite{Wilton} and references therein). Interestingly, such a re-emergence follows the dramatic decline in STIs that occurred after the appearance of the HIV in the early 1980s and the consequent widespread use of condoms. However, the introduction of the antiretroviral therapy for HIV in 1996 and a higher adoption of non-condom HIV risk-reduction strategies led to a decreased condom use and the re-emergence of STIs in USA, Canada, and Europe \cite{Fenton,Wilton}.
In this context, behavioral interventions remain an important tool in the global fight against STIs \cite{Task}.

For \textit{non-reactive} SAIS models \emph{without decay of awareness,}  it was shown in~\cite{Sahneh12a} that the spread of awareness causes an elevation of the
epidemic threshold.
Even if the basic reproduction number $R_0 > 1$ in the underlying model of disease transmission, an endemic equilibrium only appears once this elevated
 threshold  is surpassed. However, this result requires the assumption that awareness will not decay over time. In \cite{Juher14}, it was proved that this elevated
epidemic threshold disappears if one assumes that awareness will decay over time. Under this assumption, when $R_0 > 1$, awareness may drive the endemic equilibrium to very low levels of disease prevalence, but may not eliminate  it or change its stability. Here we show that \emph{reactive} SAIS models with awareness decay in some cases permit an elevated
epidemic threshold  so that the endemic equilibrium will disappear and the disease will be driven towards extinction from almost all initial states, even when $R_0 > 1$.

Will a behavioral response that is induced by awareness  prevent, all by itself, future flare-ups from low endemic levels? When awareness is assumed to be permanent and demography is not considered, as in \cite{Sahneh12a,Sahneh11}, the answer is obviously yes. But it is less clear what to expect  when awareness decays over time. In the context of ODE models the question translates into one about the existence of sustained oscillations with a significant amplitude. In Section~\ref{SAIS-Non} we show that such oscillations are ruled out in SAIS models. This result  applies to both reactive and non-reactive SAIS models and holds even if we allow more general functional responses rather than rate constants in the models.

However, in Section~\ref{SAUIS} we  introduce a more general class of models that we call \emph{SAUIS models.} They have two distinct compartments of aware hosts who differ in their willingness to alert other susceptible hosts. Thus these models embody the \emph{degradation of  quality of information} as it is transmitted from one individual to another. Our approach for modeling this phenomenon adopts,  albeit in greatly simplified form, an idea that was introduced in  \cite{ABCN} and adopted in \cite{Funk09} for an epidemic context.   We  show that sustained oscillations can occur in SAUIS models, even if all rate coefficients are constants. Also, while it seems quite plausible that oscillations could be induced by a time lag between actual prevalence and available information about it, this mechanism is deliberately ruled out by the way we set up our models.  Thus our results clearly demonstrate that degradation of information during transmission processes  can be the driving mechanism  for  the existence of periodic waves of infection, supporting the claim in \cite{ABCN} that such a degradation can reveal important qualitative and quantitative effects.

The remainder of the paper is organized as follows: In Section~\ref{SAIS-Non} we formally define reactive SAIS models and prove the results about these models that were described above. In Section~\ref{SAUIS} we formally define reactive SAUIS models and examine some of their basic properties.   In particular, we present numerical explorations both of the dynamics in SAUIS models and of regions of the parameter space that are bounded by Hopf bifurcation points. These results reveal intricate possibilities for the dynamics in SAUIS models.

In Section \ref{sec:Discussion} we briefly review some related models of behavioral epidemiology that predict oscillations and discuss how they differ from ours. We also discuss some possible implications for public health policy and directions of future work.

%%%%%%%%%%%%%%%%%%%%%%%%%%%%%%%%%%%%%%%%%%%%%%

\section{Reactive SAIS models}\label{SAIS-Non}

\subsection{The model}

An SAIS model has three compartments: S (susceptible), A (aware) and I (infectious). Susceptible hosts can move to the A-compartment or to the I-compartment, aware hosts can move to the S-compartment due to awareness decay or to the I-compartment due to infection (albeit at a lower rate than susceptible hosts). Upon recovery,  infectious hosts can move either to the A-compartment or to the S-compartment.

As was mentioned in the introduction, we will consider awareness as a state of knowledge about the prevalence of the disease that will both induce a behavioral response in the given host and that this host is willing to transmit to other hosts.  So, it is natural  to assume that awareness can be passed on from aware to unaware individuals like a contagious disease~\cite{EPCH}. However, while the force of infection in transmission of actual diseases can usually be assumed to be a linear function of the prevalence, the rate at which susceptible individuals become aware usually will show more pronounced nonlinearities.  On the one hand, there will be a saturation effect arising from overexposure to information when the proportion of infectious hosts is very high.   On the other hand, when the prevalence of the disease is very low, a careless attitude may prevail that renders the channels of transmitting awareness ineffective. Details may significantly vary between different populations and diseases.  In order to allow for maximum flexibility in incorporating these effects, we will investigate models where parameters for transmission of awareness are functions of the disease prevalence rather than rate constants.  The sources of nonlinearities mentioned above suggest that these parameters may exhibit near switch-like behavior.

The proportions of hosts in the S-, A-, and I- compartments will be denoted by $s, a, i$ respectively.
The rates of change of these fractions are governed by the following ODE model:

\begin{equation}\label{rSAISa}
\begin{split}
\frac{da}{dt} &=  \alpha_i(i) \, s \, i + \alpha_a(i) \,s \, a + p(i)\, \delta \, i -  \beta_a \, a \, i - \delta_a(i) \, a, \\
\frac{di}{dt} &=  (\beta \, s  + \beta_a\, a - \delta) \, i, \qquad s+a+i = 1.
\end{split}
\end{equation}

\noindent
Here we assume that  $\alpha_i(i) $,  $\alpha_a(i)$, $p(i)$, $\delta_a(i)$ are nonnegative differentiable functions in $[0,1]$, $p(i) \le 1$,  $\delta_a(0) > 0$, and $\beta, \beta_a, \delta$ are constants such that $0 \leq \beta_a < \beta$ and $\delta>0$. Moreover, for $i > 0$ we assume that $\alpha_a(i) > 0$ and $\alpha_i(i) + p(i) > 0$.

The term  $\alpha_i(i)i$ represents the rate at which a susceptible host becomes aware due to \emph{direct information} about the disease prevalence.  As $\alpha_i(i)$ may depend on $i$, the factor $i$ is strictly speaking redundant in this term. But its inclusion simplifies some calculations.   Also, inclusion of the factor $i$ makes  the similarity with disease transmission more explicit.  If  $\alpha_i(i)$ is not constant, one can think of the process of obtaining direct information as encountering at least one infectious host and then seeking or re-interpreting independent information about the overall disease prevalence.  Thus it seems plausible to assume that $\alpha_i(i)$ is low when infectious hosts are observed so rarely that they are not considered indicative of an outbreak, steeply increases around a critical level of disease prevalence, and then levels off.

Similarly, the term $\alpha_a(i)a$ represents the rate at which susceptible hosts become aware due to a contact with an aware host during which the latter transmits information about the disease. The assumption that $\alpha_a(i) > 0$ is the one that really distinguishes our models from  \emph{non-reactive SAIS models.} The latter can be obtained by simply removing the term~$\alpha_a(i) \,s \, a$ from~\eqref{rSAISa}. The previously published versions of non-reactive SAIS models also do not include a term~$p(i)\, \delta \, i$.

In view of the above discussion we tend to think of~$\alpha_i(i)$ and~$\alpha_a(i)$  as increasing in a sigmoid-like fashion from~0 or near 0 when $i = 0$ to near a saturation constant when $i = 1$, but these properties are not needed for the results of this section. While one can argue that $\alpha_a(0)$ should be zero so that awareness will not spread in the absence of an outbreak, we also allow for a minimum level of creation of awareness ($\alpha_a(0) > 0$) reflecting, for instance, the spread of rumors and beliefs, or a propensity to become aware because of previous experiences with the disease, even in the absence of current empirical evidence.

The parameter~$p(i)$ can be interpreted as the probability that an infectious host will move to the A-compartment as a result of \emph{direct experience} of the disease. With probability $1-p(i)$ the host will remain oblivious of the dangers posed by the outbreak and will move back to the S-compartment.

The term $\delta_a(i)$ represents the decay of awareness. It could be  a constant or any other positive differentiable function, but in realistic models higher disease prevalence should result in slower awareness decay so that $\delta_a(i)$ will be a nonincreasing function of $i$.

The inequality $\beta_a < \beta$ embodies the assumption that awareness will lead to adoption of a behavioral response that decreases the rate at which hosts contract the infection.

{\lemma  Assume  $\alpha_i(i), \alpha_a(i), p(i), \delta_i(i), \beta$ satisfy the conditions that were spelled out below \eqref{rSAISa}.
Then the region $\Omega = \{ (a,i) \in \mathbb{R}^2 \, | \, 0 \le \,  a + i \le 1, a \in [0,1] \}$ is forward-invariant. \label{invariance}}

\noindent \textit{Proof.}  By direct inspection of the system we see that $\left.(da/dt)\right|_{a=0} = \alpha_i(i) (1-i) i + p(i)\delta i \ge 0$ for $0 \le i \le 1$, $\left. (di/dt)\right|_{i=0} = 0$, and $\left. (d(a+i)/dt)\right|_{a+i=1} = -(1 - p(i))\delta i - \delta_a(i)a  \leq 0$.
$\Box$

\subsection{Nullclines and equilibria}

By solving $di/dt=0$ we find two  parts of the $i$-nullcline. The first one is given by the horizontal axis $i=0$ and the second one is the straight line
\begin{equation}\label{eqn:i-null-(a)}
i(a) = 1 -\frac{\delta}{\beta} - \left( 1 -\frac{\beta_a}{\beta} \right) a,
\end{equation}
which has a  slope between $-1$ and 0 under the assumption $\beta_a < \beta$.  It intersects the horizontal axis~$i=0$ at the point
$$a = \frac{\beta - \delta}{\beta - \beta_a}.$$

By solving $da/dt=0$ we find the $a$-nullcline. The point $(0,0)$ always satisfies this equation. For $i > 0$ we have assumed $\alpha_a(i) > 0$ and the other part of this nullcline is  implicitly defined by the following equation in the variables $i$ and $a$:
\begin{equation}\label{eqn:a-null-implicit}
a^2 - \left( 1 -  i - \frac{(\alpha_i(i) + \beta_a) i + \delta_a(i)}{\alpha_a(i)} \right) a - \frac{\alpha_i(i)}{\alpha_a(i)} \,i \, (1-i)  - \frac{p(i)\delta}{\alpha_a(i)}i = 0.
\end{equation}
Since we have assumed $\alpha_i(i) + p(i) > 0$ for $i > 0$, the positive branch of the $a$-nullcline is given by the graph of the following function~$a(i)$ on~$[0,1]$:

{\small
\begin{eqnarray}\label{eqn:a-null-pos}
a(i) & = & \frac{1}{2} \, \left( 1 - i - \frac{(\alpha_i(i) + \beta_a) i + \delta_a(i)}{\alpha_a(i)} \right. \\
& & \left.
+ \   \sqrt{\left(  1 - i - \frac{(\alpha_i(i) + \beta_a) i + \delta_a(i)}{\alpha_a(i)} \right)^2 + 4 \frac{\alpha_i(i)}{\alpha_a(i)} \, i \,(1-i) + \frac{4p(i) \delta}{\alpha_a(i)}i } \right).\notag
\end{eqnarray}
}

\noindent
Note that  $a(1) \geq 0$.  If $p(1) = 0$, then $a(1) = 0$. If $p(1) > 0$, the last term under the root in~\eqref{eqn:a-null-pos} will be positive so that $a(1) > 0$ and
the curve $a(i)$ will enter~$\Omega$ only at some point~$(a,i)$ with~$i < 1$. Similarly, $a(0) = 1 - \delta_a(0)/\alpha_a(0)$ if $\delta_a(0) < \alpha_a(0)$, while $a(0) = 0$ if $\delta_a(0) \ge \alpha_a(0)$.  Moreover, $a(i)$ is continuous and takes on positive values for all~$i \in (0,1)$.

The system has three possible types of equilibria in the first quadrant, namely,
$$
P_1=(0,0), \quad P_2 = \left( 1-\frac{\delta_a(0)}{\alpha_a(0)}, 0 \right), \quad P_3 = (a^*, i^*),
$$
where $(a^*, i^*)$ denotes an  equilibrium with $i^* > 0$. Since for $i^* > 0$ we have $\alpha_i(i^*) + p(i^*) > 0$ by  assumption, any endemic equilibrium must necessarily lie in the interior of~$\Omega$. $P_2$ lies in $\Omega$ provided that $\delta_a(0) < \alpha_a(0)$.

\begin{figure}
\begin{center}
\begin{tabular}{c}
\includegraphics[scale=0.4]{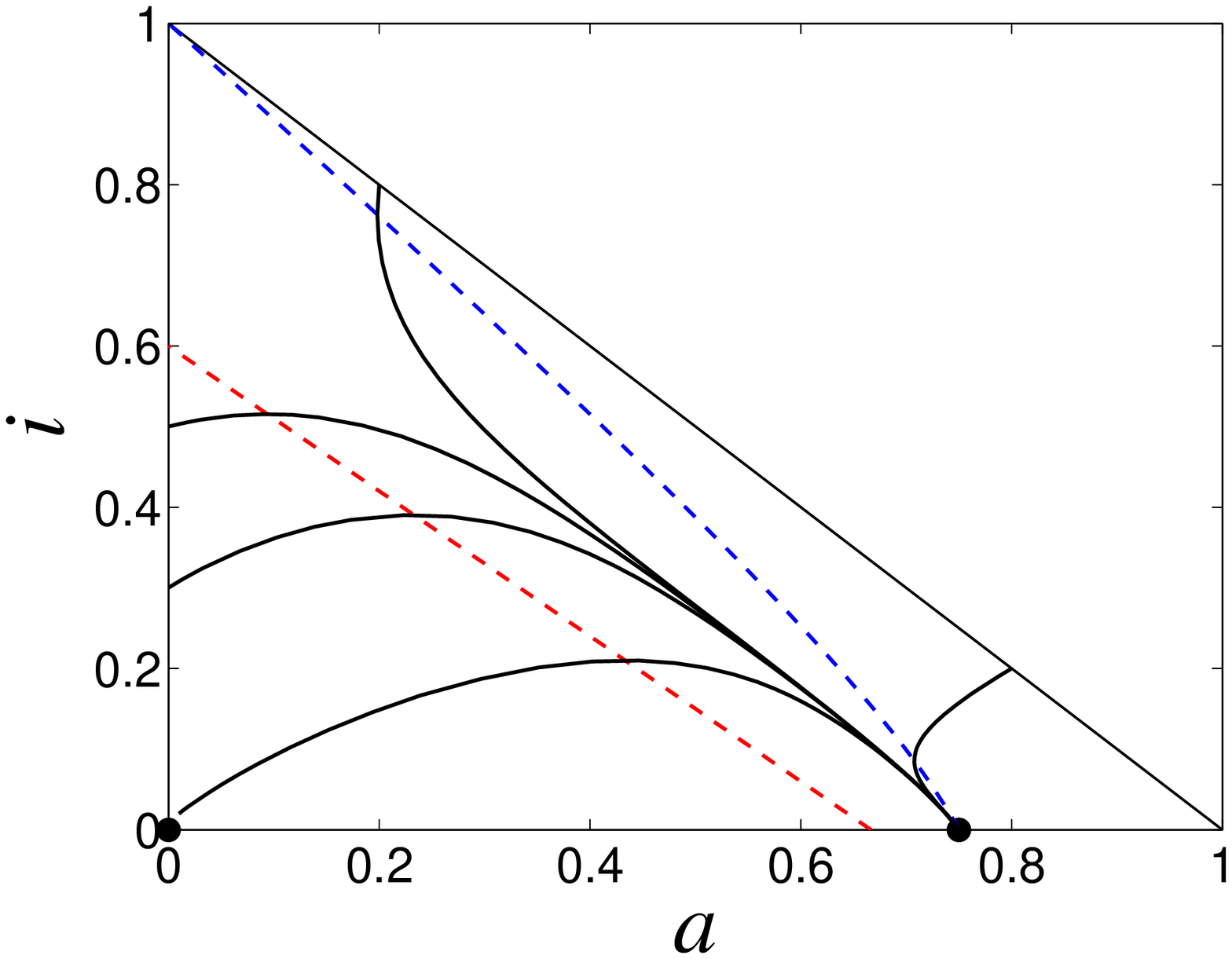}
\\
\includegraphics[scale=0.4]{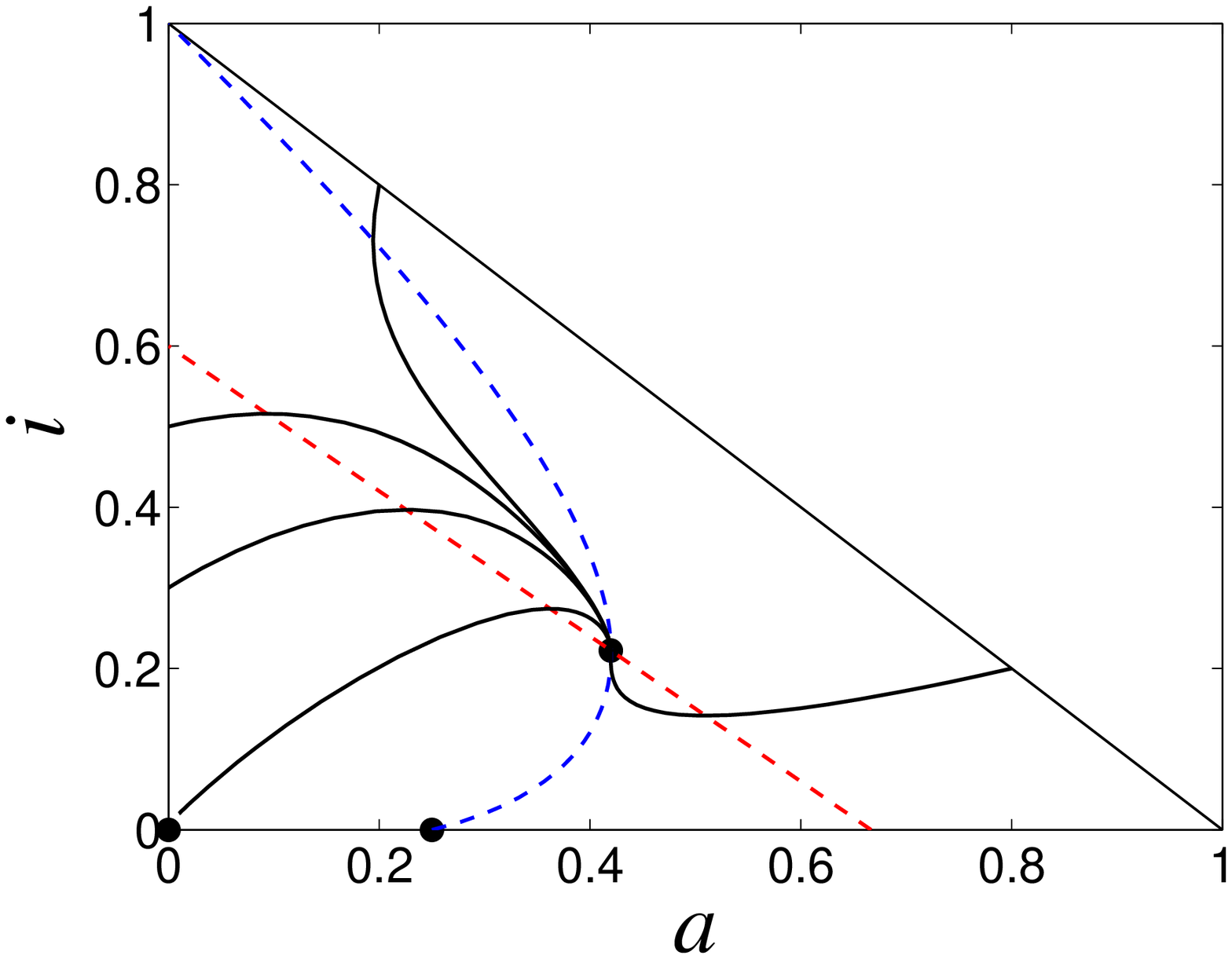}
\\
\includegraphics[scale=0.4]{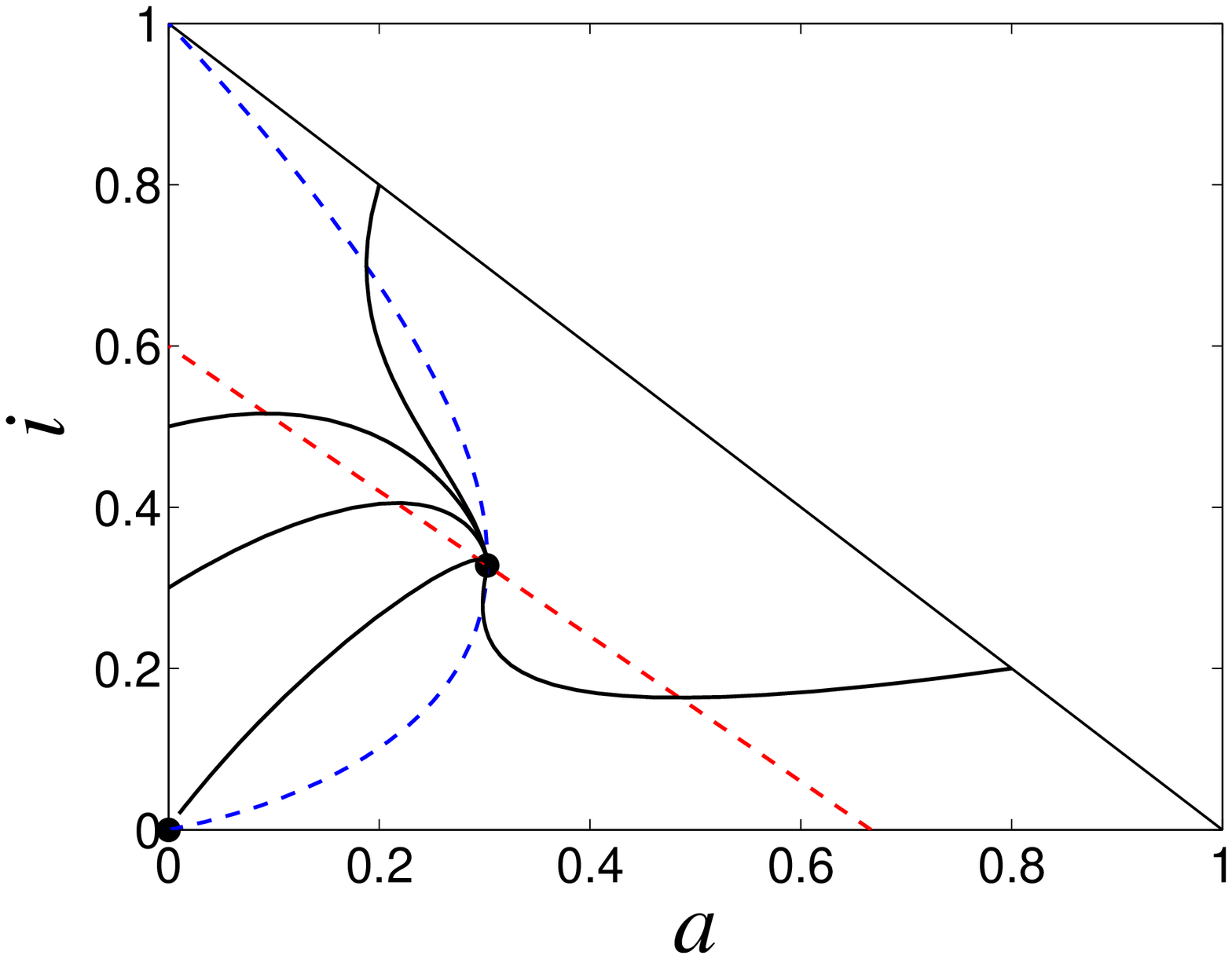}
\end{tabular}
\caption{Phase portrait of the reactive SAIS with alert rates $\alpha_a(i)=\alpha^0_a \, (i+1)$ and  $\alpha_i(i)=\alpha^0_i \, (i+1)$, and decay rate $\delta_a(i)=\delta_a^0/(1+i)$ for different values of $\delta_a^0$ showing three possible configurations of equilibria when $R_0 >1$ (top: $\delta_a^0=1$, middle: $\delta_a^0=3$, bottom: $\delta_a^0=5$). Parameters: $p(i) = 0$, $\delta=4$, $\beta=10$, $\beta_a=1$, and $\alpha_a^0=\alpha_i^0=4$. Note that $\alpha_a(0)  > 0$  allows the existence of a second equilibrium on the $a$-axis for small enough values of $\delta_a^0$.
\label{Portraits-rSAIS}}
\end{center}
\end{figure}

Now consider the inequality
\begin{equation}\label{eqn:P3-cond}
\frac{\beta-\delta}{\beta-\beta_a} > 1 - \frac{\delta_a(0)}{\alpha_a(0)}.
\end{equation}
By sketching  the nullclines in the $a$-$i$ plane, one can see that~\eqref{eqn:P3-cond} is a sufficient condition
for the existence of at least one endemic equilibrium when $\beta > \delta > \beta_a$, because the function~$a(i)$ is continuous  and satisfies $a(1) \geq 0$. It will always contain a point on the boundary of~$\Omega$ with $a+i = 1$.  Thus~\eqref{eqn:P3-cond} guarantees that it must intersect the part of the $i$-nullcline given by~\eqref{eqn:i-null-(a)}, which is a straight line such that $i(0) < 1$ with a slope larger than $-1$  and $a$-intercept $\frac{\beta-\delta}{\beta-\beta_a} < 1$ (see Figure \ref{Portraits-rSAIS}). The condition $\delta > \beta_a$ can be relaxed when all the coefficients in the model are constants; see Lemma~\ref{lem:unique}(a) and its proof.

Alas, our model assumptions do not rule out multiple endemic equilibria~$P_3$.  Neither is~\eqref{eqn:P3-cond} necessary for the existence of endemic equilibria. Under a number of fairly natural conditions, uniqueness of~$P_3$ is guaranteed and~\eqref{eqn:P3-cond} is necessary for its existence; see Lemma~\ref{lem:unique} below. However, as the main results of this section can be derived under our most general assumptions, we will not impose any of these conditions from the outset. We want to mention though that the case of constant rate functions that is covered by points~(a1) and~(b1) of Lemma~\ref{lem:unique} is the only one that has been studied in the prior literature  on nonreactive SAIS models, and is also the one that directly corresponds to our work in the next section.

\medskip

\begin{lemma}\label{lem:unique}
Assume  $\alpha_i(i), \alpha_a(i), p(i), \delta_i(i), \beta$ satisfy the conditions that were spelled out below \eqref{rSAISa}, and that $\beta > \delta$. Then:

\begin{itemize}
\item[(a)] Under any of the following assumptions, when~\eqref{eqn:P3-cond} holds, there exists exactly one interior equilibrium~$P_3$:

\smallskip

\begin{itemize}
\item[(a1)] The functions $\alpha_i(i) = \alpha_i, \alpha_a(i) = \alpha_a, p(i) = p$ and $\delta_a(i) = \delta_a$ are all constant.

\smallskip

\item[(a2)] The functions $\alpha_i(i), \alpha_a(i), p(i)$ are nondecreasing, $\delta \geq \beta_a$, the function~$\delta_a(i)$ is nonincreasing,  and  $\alpha_a(0) > \beta - \beta_a.$

\smallskip

\item[(a3)] The functions $\alpha_i(i), \alpha_a(i), p(i)$ are nondecreasing, $\delta \geq \beta_a$, and the function~$\delta_a(i)$ decreases steeply enough so that  $\beta_a \leq -  \delta_a'(i)$ for all~$i \in [0,1]$.
\end{itemize}

\smallskip

\item[(b)] If the inequality in~\eqref{eqn:P3-cond} is reversed, then any of the following conditions precludes the existence of an interior equilibrium.

\smallskip

\begin{itemize}
\item[(b1)] The functions $\alpha_i(i) = \alpha_i, \alpha_a(i) = \alpha_a, p(i) = p$ and $\delta_a(i) = \delta_a$ are all constant and
\begin{equation}\label{eqn:b1-cond}
\frac{\alpha_a - \delta_a}{\alpha_a + \beta_a} \geq 1 - \frac{\delta}{\beta}.
\end{equation}
\smallskip

\item[(b2)] The assumptions of~(a2)  hold.
\smallskip

\item[(b3)] The assumptions of~(a3)  hold.
\end{itemize}
\end{itemize}
\end{lemma}

\begin{figure}
\begin{center}
\begin{tabular}{cc}
\hspace{-1cm}
\includegraphics[scale=0.36]{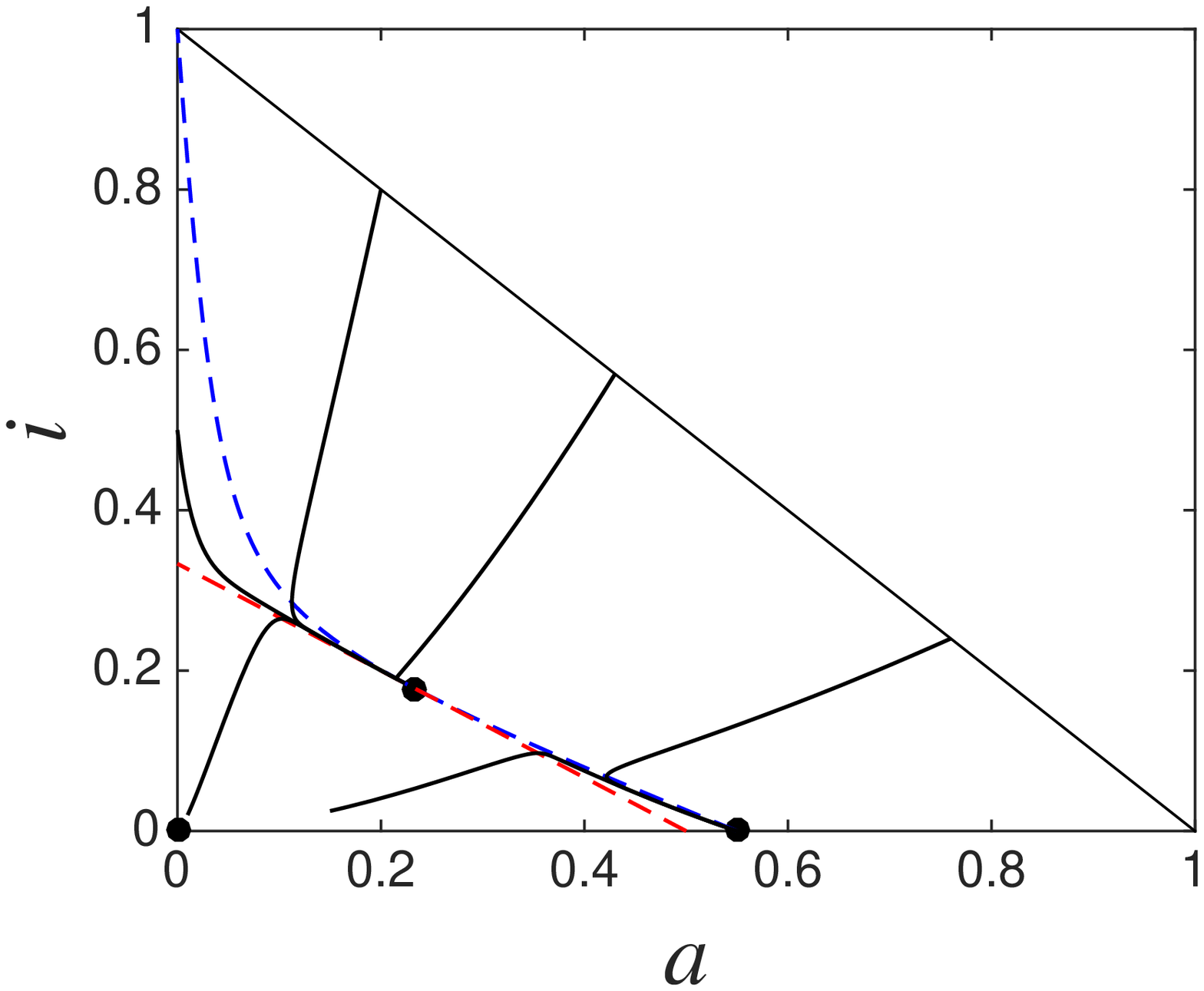}
&
\hspace{-0.75cm}
\includegraphics[scale=0.36]{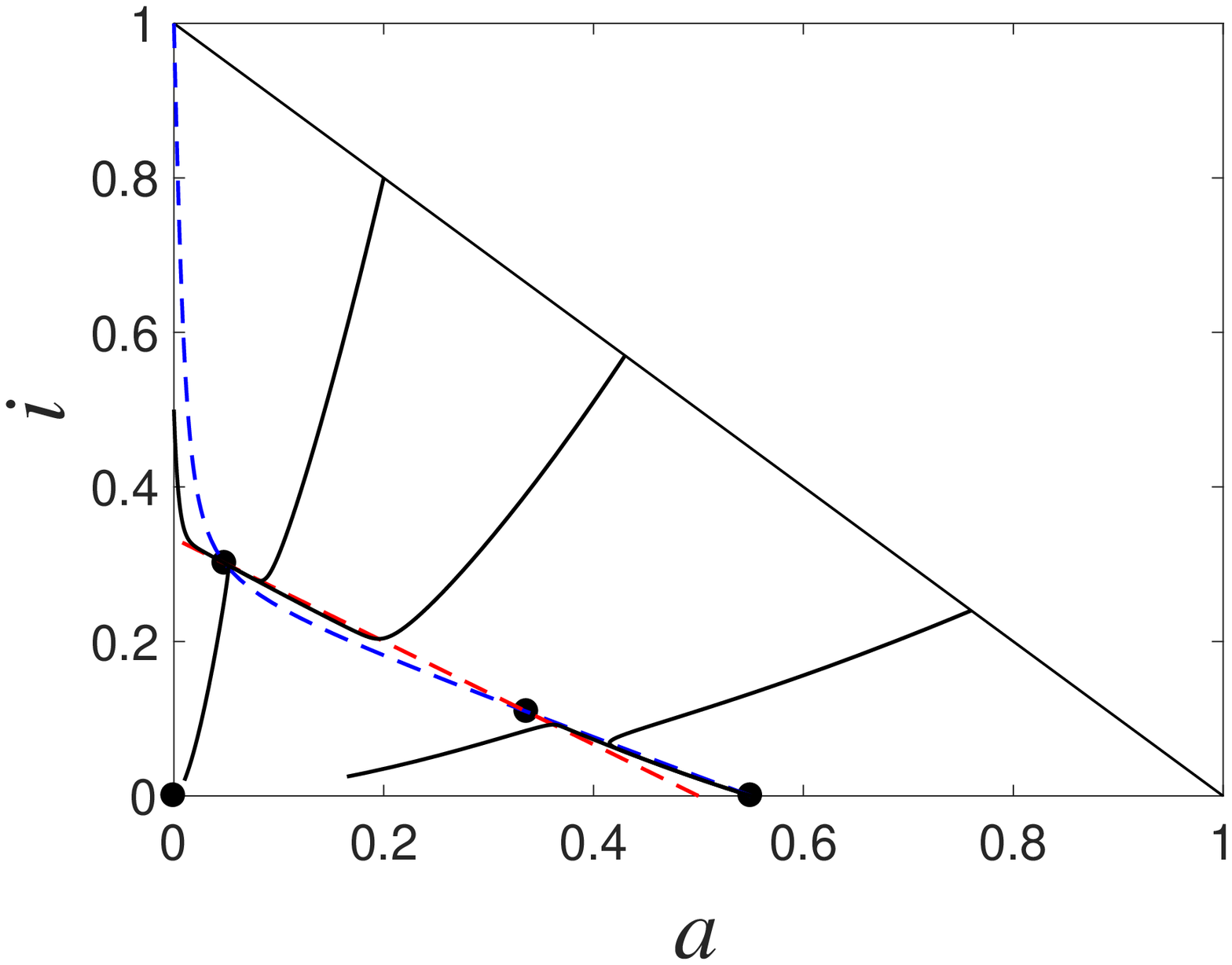}
\end{tabular}
\caption{Phase portrait of the reactive SAIS with constant rates showing the existence of two interior equilibria (right) after a saddle-node bifurcation (left) using $\alpha_i$ as a tuning parameter. Parameters: $p=0$, $\beta=6$, $\delta=4$, $\beta_a=2$, $\delta_a=0.9$,
$\alpha_a=2$, and $\alpha_i=0.05$ (right) and $\alpha_i=0.1733500838578$ (left).
\label{Fig:Saddle-node}}
\end{center}
\end{figure}

The proof of this lemma is included in the appendix. In part~(b1), without an assumption like~\eqref{eqn:b1-cond}, saddle-node bifurcations that lead to multiple endemic equilibria are possible; see Figure~\ref{Fig:Saddle-node}.

\bigskip

If we define the basic reproduction numbers of the disease and awareness as
$$
R_0 := \beta/\delta \qquad \mbox{and} \qquad R_0^{\,a} := \alpha_a(0)/\delta_a(0),
$$
we can interpret intuitively the conditions for the existence of these equilibria. The disease-and-awareness-free equilibrium $P_1$ always exists. The equilibrium $P_2$ is also disease-free, but has a positive fraction of aware hosts. It exists if, and only if,  $R_0^{\,a} > 1$, that is, if  in  a large and otherwise susceptible population with one aware host,  awareness will on average increase. The condition $R_0 > 1$ is necessary and, when $\delta > \beta_a$, condition~\eqref{eqn:P3-cond} is sufficient for the existence of an endemic equilibrium~$P_3$.  Note that condition~\eqref{eqn:P3-cond} holds if the disease spreads faster than awareness in the early stages of an outbreak, i.e., if $R_0 > \max\{1, R_0^a\}$.  In addition, it also holds when $R_0^{\,a} > R_0 > 1$ and $\beta_a$ is close enough to $\delta$, i.e., when the reduction of the transmission probability  due to awareness is not significant enough.

\medskip

The Jacobian  matrix of system \eqref{rSAISa} is

\begin{equation*}
J = \left( \begin{array}{cc}
\alpha_a(i) (s\! -\!  a) \! - \! \alpha_i(i)  i \! - \!\beta_a i \! -\! \delta_a(i)\ \ & \ \ URC\\
- (\beta - \beta_a)  i  & \beta  s + \beta_a  a - \beta  i - \delta
\end{array}
\right),
\end{equation*}
where $URC = (\alpha_i(i)i)'s \!-\! \alpha_i(i)i \!+ \!(\alpha'_a(i) s \!- \!\alpha_a(i) \!- \!\beta_a \!- \!\delta'_a(i))  a \!+ \!(p(i) \delta i)'$
and $s=1-a-i$.

The eigenvalues of $J$  at $P_1$ are
$$\lambda_1(P_1) = \alpha_a(0) - \delta_a(0)\qquad  \mbox{and} \qquad \lambda_2(P_1) = \beta - \delta,$$
 which shows that the necessary conditions for the existence of $P_2$ and $P_3$ imply instability of $P_1$.

At $P_2$, the eigenvalues are
$$\lambda_1(P_2) =  \delta_a(0) - \alpha_a(0) \qquad \mbox{and}
\qquad
\lambda_2(P_2) = \beta - \delta - (\beta - \beta_a) \left(1 - \frac{\delta_a(0)}{\alpha_a(0)} \right).
$$
Note that $\lambda_1(P_2) = - \lambda_1(P_1)$ and that $\lambda_2(P_2) > 0$ if and only if condition~\eqref{eqn:P3-cond} holds. The former implies that if $R_0^a > 1$, then $P_1$ is unstable and $P_2$ attracts any trajectory on the $a$-axis;  exactly as one would expect from an SIS-like behavior of awareness.

\subsection{Dynamics}

{\lemma  Assume  $\alpha_i(i), \alpha_a(i), p(i), \delta_i(i), \beta$ satisfy the conditions that were spelled out below
\eqref{rSAISa}. Then the system \eqref{rSAISa} has no closed orbits inside $\Omega$. \label{NCO}}

\noindent \textit{Proof.}    Let $f_1(a,i)$ and $f_2(a,i)$ denote the functions on the  right-hand side of the system. The vector field $\displaystyle (F_1(a,i), F_2(a,i)) = \left( \frac{1}{a\,i}  f_1(a,i), \frac{1}{a\,i}  f_2(a,i) \right)$ is $C^1$ in the interior of $\Omega$, and its divergence is given by
$$
\frac{\partial}{\partial a} F_1(a,i) + \frac{\partial}{\partial i} F_2(a,i) = - \frac{\alpha_i(i)}{a} \left(1 + \frac{s}{a} \right) -  \frac{\alpha_a(i)}{i} - \frac{p(i) \delta}{a^2} - \frac{\beta}{a}   <  0
$$
for all $(a,i)$ in the interior of $\Omega$. So, the divergence does not change sign and does not take the value $0$.  Therefore,  Dulac's criterion of nonexistence of periodic orbits \cite{Perko}  precludes the existence of a closed orbit lying entirely in~$\Omega$.
$\Box$

\medskip

The next theorem sums up the dynamics of system.

{\theorem Assume  $\alpha_i(i), \alpha_a(i), p(i), \delta_i(i), \beta$ satisfy the conditions that were spelled out below
\eqref{rSAISa}. Then the global behavior of the solutions of the system \eqref{rSAISa} depends as follows on the remaining parameters:

\begin{enumerate}
\item[(i)] If $R_0 \leq 1$ and $R^a_0 \leq 1$, then $P_1$ is the only equilibrium point and is globally asymptotically stable.

\item[(ii)] If $R_0 \leq 1 < R^a_0$, then $P_1$ and $P_2$ are the only equilibrium points. $P_2$ is globally asymptotically stable on $\Omega \backslash \{P_1\}$. When
    $R_0 < 1$, then $P_1$ is a saddle point.

\item[(iii)] If $R^a_0 \leq 1 < R_0$, then no equilibrium~$P_2 \neq P_1$ exists in~$\Omega$.  When
   $R^a_0 < 1$, then $P_1$ is a saddle point. Each trajectory that starts with $i(0) > 0$ will eventually approach an endemic equilibrium of type~$P_3$.

\item[(iv)] If $R_0 > 1$ and $R^a_0 > 1$, then $P_1$ is an unstable point and system \eqref{rSAISa} has also the equilibrium $P_2$. If  $\lambda_2(P_2) < 0$,
then~$P_2$ is locally asymptotically stable.

\item[(v)] If instead $\lambda_2(P_2) > 0$, then system \eqref{rSAISa} has also at least one equilibrium $P_3$, with $P_2$ being a saddle point.  Each trajectory that starts with $i(0) > 0$ will eventually approach an endemic equilibrium of type~$P_3$.
\end{enumerate}

Moreover, when any of the conditions of Lemma~\ref{lem:unique}(a) are satisfied,
then the endemic equilibria in point~(iii) and~(v) are guaranteed to be unique. Similarly, when any of the conditions of Lemma~\ref{lem:unique}(b) are satisfied,
then under the assumptions of point~(iv) we can conclude that~$P_2$ is globally asymptotically stable on $\Omega \backslash \{P_1\}$.
\label{SAIS-thm}}

\noindent \textit{Proof.}  Recall that $R_0 = \frac{\beta}{\delta}$ and $R^a_0 = \frac{\alpha_a(0)}{\delta_a(0)}$.

For  part~(i), when $\beta \leq \delta$, then the  $i$-nullcline has no points with $s < 1$    and
$P_1$ is the only equilibrium point of \eqref{rSAISa}.
 From Lemmas~\ref{invariance} and~\ref{NCO}, together with the Poincar\'e-Bendixson theorem, it follows that $P_1$ is globally asymptotically stable.

For  part~(ii), if  $\alpha_a(0) > \delta_a(0)$, the $a$-nullcline intersects the $a$-axis at $P_2$, and $P_3$ is ruled out as in case~(i). So $P_1$ and $P_2$ are the only equilibria of the system. As we have $\lambda_1(P_1) > 0$ and $\lambda_2(P_1) = \beta - \delta \leq  0$, it follows that  $P_1$ is a saddle point when the inequality $\beta < \delta$ is strict. Moreover, as  $(\alpha_i(0) i)' \geq 0$, the  first row of the Jacobian at $P_1$ has two nonnegative entries. Thus the eigenvector with eigenvalue $\lambda_1(P_1)$ cannot intersect the interior of $\Omega$.
From the same facts that we used in part~(i)
it follows that $P_2$ is globally asymptotically stable on $\Omega \backslash \{P_1\}$.

Under the assumptions of (iii), we have $\lambda_1(P_1) = \alpha_a(0) - \delta_a(0) \leq  0$ and $\lambda_2(P_1) > 0$, and it follows that  $P_1$ is a saddle point when the inequality $\alpha_a(0) < \delta_a(0)$ is strict. Moreover, the eigenvector with non-positive eigenvalue lies on the line~$i = 0$.
Thus
at least one endemic equilibrium~$P_3$ must exist, and each trajectory that starts with~$i(0) > 0$ must eventually approach such an equilibrium.

Under the assumptions of  (iv) and (v), the $a$-nullcline intersects the horizontal axis at $a(0) > 0$ and $P_2$ is always an equilibrium, while $P_1$ is unstable.
Recall that the $i$-nullcline intersects the $a$-axis at
$(\beta - \delta)/(\beta - \beta_a)$ so that  the sign of $\lambda_2(P_2)$ determines whether this intersection occurs to the left or to the right of $P_2$.  Note that this remains true even when this intersection occurs outside of~$\Omega$.
The former occurs in case~(iv) where $P_2$ is locally asymptotically stable.
The latter occurs in case (v), where $P_2$ is a saddle point with the stable direction on the line~$i = 0$, while $P_1$ is repelling.
It follows that at least one endemic equilibrium~$P_3$ must exist, and each trajectory that starts with~$i(0) > 0$ must eventually approach such an equilibrium.
$\Box$

\medskip

For $R_0^{a} \ge 1$ (so that $P_2 \ge 0$), Theorem \ref{SAIS-thm} shows that  the reactive SAIS-model predicts an elevated epidemic threshold which is given by $\lambda_2(P_2)=0$. This threshold can be written by changing the inequality in~\eqref{eqn:P3-cond} to an equality.
Following the suggestion of one of the reviewers, let us  define
\begin{equation*}
R_0^d = \frac{\beta_a}{\delta}
\end{equation*}
as the basic reproduction number of aware individuals
that would apply to a population consisting entirely of aware hosts.
After dividing numerator and denominator of the left-hand side of \eqref{eqn:P3-cond} by $\delta$ and using the definitions of $R_0$ and $R_0^a$,
inequality~\eqref{eqn:P3-cond} now reads
$$
\frac{R_0-1}{R_0-R_0^d} > 1 - \frac{1}{R_0^a} = \frac{R_0^a - 1}{R_0^a}.
$$
After passing all terms to the left-hand side of the inequality, multiplying  by the positive denominators, canceling $R_0R_0^a$, and then adding 1 to both sides,   this can be written equivalently in the usual format for epidemic thresholds as
\begin{equation}\label{eqn:new-threshold}
R_0  + (R_0^a - 1)(R_0^d-1) > 1.
\end{equation}

When \eqref{eqn:new-threshold} holds and $R_0^a \ge 1$,  the disease can always invade the population regardless of how much awareness is initially present, and trajectories that start at an endemic state will eventually approach an endemic equilibrium. In particular, when $R_0^a \geq 1$ and $R_0^d \geq 1$, then~\eqref{eqn:new-threshold} holds, as  $R_0 = \beta/\delta > R_0^d$ under  the assumptions of our model, and the epidemic will spread. On the other hand, if $R_0^a > 1$ but $R_0^d < 1$ (awareness significantly reduces disease transmission), then having $R_0 > 1$ is not enough to guarantee the spread of the epidemic.

When the inequality in~\eqref{eqn:new-threshold} is reversed and still $R_0^a \ge 1$, then the disease will not be able to invade a population that has had some prior exposure to awareness, by whatever means, and has reached a state in the basin of attraction of~$P_2$. However, it might still be possible for the disease to persist outside of this basin of attraction; see the example of Figure~\ref{Fig:Saddle-node}, where the left-hand side of~\eqref{eqn:new-threshold} evaluates to~$0.888$. Therefore, when $R_0^a > 1$ and $R_0^d < 1$, which is feasible in a region of parameter space for our model,  \eqref{eqn:new-threshold} defines an elevated epidemic threshold. If in addition any of conditions  of Lemma~\ref{lem:unique}(b) hold, then the disease will always be driven to extinction. Under our most general assumptions,  the basin of attraction of of $P_2$ will include an open ball around~$P_2$ as well as the line segment $L$ that consists of all conditions with $i = 0$ and $a > 0$.
It follows that this basin of attraction of~$P_2$ must contain an open neighborhood of~$L$. Thus in the limiting case of infinite population size where introduction of a single index case is treated as an initial condition with an infinitesimally small positive $i(0)$, any prior introduction of awareness ($a(0) > 0$) would be sufficient to guarantee that the disease dies out before reaching endemic proportions. This conclusion may not be valid under our general assumptions for small finite populations, but exposure to awareness at a time that is sufficiently early relative to invasion of the disease would still guarantee that it will be driven to extinction under the assumptions of Theorem~\ref{SAIS-thm}(iv).

It may be of interest to point out that in the limiting case $\beta_a = 0$ the behavioral response that is triggered by awareness confers perfect immunity to becoming infectious. This is quite similar to the assumption that is often made about vaccinations. In this case the inequality in~\eqref{eqn:P3-cond} will be reversed if $R_0 < R_0^{a}$. Under these conditions, if all rate functions satisfy the monotonicity conditions of Lemma~\ref{lem:unique}(b3), then any trajectory that starts in $\Omega \backslash \{P_1\}$  will approach the equilibrium $P_2$ with a proportion of $1 - 1/R_0^{a}$ aware hosts, which happens to be exactly the herd immunity threshold for vaccinations that confer perfect protection for diseases with basic reproduction number~$R_0^{a}$.

%%%%%%%%%%%%%%%%%%%%%%%%%%%%%%%%%%%%%%%%%%%%%%%%%%

\section{SAUIS models}\label{SAUIS}

\subsection{The model}

SAIS models ignore the degradation of information quality as it is transmitted from one individual to another.  According to these models, aware individuals would always create aware hosts with the same degree of responsiveness, which seems unrealistic. The following approach  to modeling degradation of information was introduced  \cite{ABCN} and adopted in \cite{Funk09} for an epidemic context:  We assume that direct information about the disease prevalence induces awareness of the risk of infection and maximum observance of protective measures,  while subsequent awareness transmission decreases its impact on an individual's reaction by a constant decay factor. Following this idea, but simplifying it in order to get a manageable model, we will introduce a new class of individuals   whose  behavioral response  has been induced    indirectly through contacts with  aware individuals.  In contrast to aware individuals,  the latter are assumed to be unwilling to convince other people about the risk. They may also show a weaker behavioral response.  As they don't actively participate in the dissemination of awareness, we call them \textit{unwilling} individuals and let~$u$ denote their proportion in the population.
Transmission of awareness can create both aware and unwilling hosts.

We will investigate here the simplest equations describing such dynamics that are constructed in direct analogy to the reactive SAIS models of Section~\ref{SAIS-Non}:

\begin{equation}\label{eqn:SAUIS}
\begin{split}
\frac{da}{dt} & =  \alpha_i \, s \, i + \alpha_a \, s \, a + p \,\delta \, i -  \beta_a \, a \, i - \delta_a\, a, \\
\frac{du}{dt} & =  \delta_a\, a  + \alpha_u \, s \, a  + q \, \delta \, i -  \beta_u\, u \, i - \delta_u\, u,\\
\frac{di}{dt} & =  (\beta  \, s + \beta_a \, a + \beta_u \, u - \delta)\, i, \qquad s+a+u+i = 1.
\end{split}
\end{equation}
Here $\alpha_u\, a$ is the rate at which susceptible hosts become unwilling after having a contact with an aware host, and $\delta_u$ is the rate of awareness decay of the unwilling hosts.  This model implicitly assumes that aware hosts  first turn into unwilling hosts before possibly entering the susceptible compartment.
It also allows for coexistence of three behavior patterns at the time of recovery from the disease: The host will then  move into the A-compartment with probability~$p$, into the U-compartment with probability~$q$, and into the S-compartment with probability~$1-p-q$.
All other terms play the same role as the corresponding terms in the reactive SAIS model.

We assume that all rate coefficients are constants and, with the possible exception of~$\beta_a, \beta_u$, are positive, with  $0 \leq \beta_a , \beta_u < \beta$.  Similarly to the SAIS model, one could allow some of the rate coefficients to depend on~$i$; see Section~\ref{sec:Discussion} for a brief discussion of these extensions of the model. However, we specifically restrict our attention here to the case of constant rate coefficients to emphasize the point that their dependence on~$i$ is not a necessary condition for
observing periodic oscillations.

{\lemma
The region $\Omega = \{ (a,u,i) \in \mathbb{R}^3_+ \, | \, 0 \le \,  a + u + i \le 1\}$ is positively invariant.
\label{invariance2}}

\medskip

\noindent \textit{Proof.}
Direct inspection of the system \eqref{eqn:SAUIS} shows that  $\left.(da/dt)\right|_{a=0} \geq 0$  and the inequality is strict when $si > 0$.  Similarly,  $\left. (du/dt)\right|_{u=0} \geq 0$, and the inequality is strict when $a > 0$. On the other hand,  $\left. (di/dt)\right|_{i=0} = 0$. It follows that the $(a,i)$- and $(u,i)$-coordinate planes  repel the trajectories  and that the $(a,u)$-plane is invariant. So, we only need to see that trajectories cannot cross the boundary $a+u+i=1$. If $\vec{v}$ denotes the vector field defined by the right-hand side of the system, we can see this by computing  the scalar product of $\vec{v}$ on this boundary with the outward-pointing normal vector $\vec{n}=(1,1,1)$. Since $s = 0$ in this region of the boundary, we get  $\vec{v} \cdot \vec{n} = - \delta_u \, u - \delta \, i \, (1 - p - q) \leq 0$. The inequality is strict except at the point~$(1,0,0)$, where $da/dt = - du/dt < 0 = di/dt$. Therefore  the vector field on the boundary $a+u+i=1$  never points towards the exterior of $\Omega$.
$\Box$

\subsection{Possible equilibria}

The system~\eqref{eqn:SAUIS} can have up to three types of equilibria in~$\Omega$.

The first one is the disease-free, awareness-free equilibrium $P_1 = (0,0,0)$.

The second one is the disease-free equilibrium $P_2 = (a^*_0, u^*_0, 0)$. From the first line of $\eqref{eqn:SAUIS}$ we have $s^*_0 = \delta_a/\alpha_a$. In view of the equality $u^*_0 = 1 - a^*_0 - \delta_a/\alpha_a$, the second line of~\eqref{eqn:SAUIS} implies:

\begin{equation}\label{eqn:P2-location}
a^*_0 = \frac{ \delta_u \left( 1 - \frac{\delta_a}{\alpha_a} \right) }{
\delta_a \left( 1 + \frac{\alpha_u}{\alpha_a} \right) + \delta_u} ,
\quad
u^*_0 =
\left( 1 - \frac{\delta_a}{\alpha_a} \right)  \frac{\delta_a \left( 1 + \frac{\alpha_u}{\alpha_a} \right)}
{\delta_a \left( 1 + \frac{\alpha_u}{\alpha_a} \right) + \delta_u}.
\end{equation}

The third kind of possible equilibrium is an endemic equilibrium, i.e., a point $P_3=(a^*,u^*,i^*)$  of $\Omega$ with

\begin{equation} \label{i*}
i^* = 1 - \left(1-\frac{\beta_a}{\beta} \right) a^* - \left(1 - \frac{\beta_u}{\beta} \right) u^* - \frac{\delta}{\beta} > 0.
\end{equation}

As the upper panel of Figure~\ref{Fig:Transcritical} shows, there may be more than one endemic equilibrium in the interior of~$\Omega$.

\subsection{Existence and linear stability of equilibria}

\smallskip

The disease-free, awareness-free equilibrium $P_1 = (0,0,0)$ always exists.
Evaluating the Jacobian matrix of system~\eqref{eqn:SAUIS} at $P_1$ we have

\begin{equation}\label{JP1}
J(P_1)  =  \left( \begin{array}{ccc}
\alpha_a - \delta_a \ \ & \ \ 0 \ \ & \ \  \alpha_i + p \delta
\\
\delta_a + \alpha_u \ \ & \ \ - \delta_u \ \ & \ \ q \delta
\\
0  \ \ & \ \ 0 \ \ & \ \ \beta - \delta
\end{array}
\right),
\end{equation}
whose eigenvalues are

$$\lambda_1(P_1)=\alpha_a - \delta_a, \quad \lambda_2(P_1)= - \delta_u, \quad \lambda_3( P_1)= \beta - \delta.$$

So, as expected, when $\beta > \delta$, i.e., when $R_0 > 1$, this equilibrium is unstable. Moreover, as in the reactive SAIS model, when  $R^a_0 := \alpha_a/\delta_a > 1$, this equilibrium is unstable independently of the sign of $\beta - \delta$ and $P_2$ becomes biologically meaningful.

By~\eqref{eqn:P2-location}, $P_2$ exists in~$\Omega \backslash \{P_1\}$ if, and only if, $R^a_0 = \alpha_a/\delta_a > 1$.
In this case it must be in the interior of the intersection of the $a$-$u$-plane with~$\Omega$.
The Jacobian matrix of~\eqref{eqn:SAUIS} at this equilibrium is

{\small
\begin{equation}\label{JP2}
J(P_2)  =  \left( \begin{array}{ccc}
- \alpha_a\,  a^*_0 \ \,  & \ \, -\alpha_a\, a^*_0 \ \, & \ \, \alpha_i s^*_0 - (\alpha_a+\beta_a)a^*_0 + p \delta
\\
\delta_a + \alpha_u (s^*_0 - a^*_0) \ \, & \ \, -\alpha_u a^*_0 - \delta_u \ \, & \ \, -\alpha_u a^*_0 - \beta_u u^*_0 + q \delta
\\
0  \ \, & \ \, 0  \ \, & \ \, \beta s^*_0 + \beta_a a^*_0 + \beta_u u^*_0 - \delta
\end{array}
\right),
\end{equation} }

\noindent
with $s^*_0 = \delta_a/\alpha_a$.
Here we have used the observation that the expression $\alpha_a\, (s_0^* - a^*_0)  - \delta_a$ obtained from direct computation of the partial derivative in the upper left corner of~$J(P_2)$ simplifies to $-\alpha_a a^*_0$. When one computes the determinant of the submatrix formed by the intersection of the first two rows of $J(P_2)$ with its first two columns, the only negative term cancels out.  Moreover, the trace of this submatrix is negative, so that~$J(P_2)$ has two eigenvalues that are either negative or have negative real parts.

The third eigenvalue $\lambda_3(P_2)=\beta - (\beta - \beta_a) \, a^*_0 - (\beta - \beta_u) \, u^*_0 - \delta$ is negative if $\beta < \delta$ and $P_2 \in \Omega$. When $\beta > \delta$ and

\begin{equation}\label{eqn:P2lambda3+}
\frac{\beta - \beta_a}{\beta - \delta} \, a^*_0 + \frac{\beta - \beta_u}{\beta - \delta} \, u^*_0 < 1,
\end{equation}
then $\lambda_3(P_2)$ will be positive. By~\eqref{eqn:P2-location}, the values of~$a_0^*, u_0^*$ do not depend on the disease transmission parameters, and it can be seen from~\eqref{eqn:P2lambda3+} that there are large regions of the parameter space where~$\lambda_3(P_2)$ is positive and large regions where it is negative while $\beta > \delta$.

\subsection{Transcritical bifurcations}

\subsubsection{Transcritical bifurcation at $R^a_0=1$}

Assume $\beta < \delta$ so that $\lambda_3(P_1) < 0$.  As $\lambda_1(P_1)$ changes from negative to positive when the bifurcation parameter $R^a_0=\frac{\alpha_a}{\delta_a}$ increases past~1, the equilibrium $P_1$ loses its stability at the bifurcation value~1.  Simultaneously, $P_2$ enters the biologically feasible region~$\Omega$ and becomes locally asymptotically stable as explained in the previous subsection. Moreover, at the bifurcation value $P_1$ and $P_2$ coincide. Note also that the plane~$i = 0$ is invariant for the system~\eqref{eqn:SAUIS}, and when $a^*_0$ is negative, then the determinant of the submatrix formed by the intersection of the first two rows of $J(P_2)$ with its first two columns is negative.  This implies that when~$P_2$ crosses into the biologically feasible region, the equilibria~$P_1$ and~$P_2$ interchange their stability.

\subsubsection{Transcritical bifurcation at $R_0=1$}

For the analysis of the bifurcations of the endemic equilibrium $P_3$ from $P_1$ and $P_2$, we will use standard results for the existence of a transcritical bifurcation (see the criterion that is given right after Sotomayor's Theorem in \cite{Perko}).
In order to use them, we introduce some notation. Let~$\bf f$ denote the vector defined by the right-hand side of system \eqref{eqn:SAUIS} and let ${\bf f}_{\mu}$ be the vector of partial derivatives of its components $f_i$ with respect to a  bifurcation parameter $\mu$. Moreover, let $D{\bf f}_{\mu}$ be the Jacobian matrix of ${\bf f}_{\mu}$ and  let
$D^2{\bf f}({\bf y}, {\bf y})$ be the column vector with components
$\left(D^2{\bf f}({\bf y}, {\bf y}) \right)_k := \sum_{j,l} \frac{\partial^2 f_k}{\partial x_j \partial x_l} y_j y_l$,  where $\bf y$ is a vector in $\mathbb{R}^3$, $x_1=a$, $x_2=u$, and $x_3=i$. We will use ${\bf f}^{BP}, {\bf f}^{BP}_\mu$  to indicate that the relevant objects are computed at the bifurcation point.

The endemic equilibrium $P_3$ can bifurcate from $P_1$ when $\alpha_a < \delta_a$ (i.e., $R_0^a <1$). In particular, since $\lambda_3(P_1)=\beta-\delta$ is a simple eigenvalue, a bifurcation occurs for $\beta = \delta$ (i.e., at $R_0 =1$).
Taking $\beta$ as  the bifurcation parameter and evaluating the Jacobian matrix $J(P_1)$ at the bifurcation point, it follows that the row vector ${\bf u} = (0,0,1)$
 and the column vector
${\bf v} = (1, (\delta_a + \alpha_u + \frac{q\delta(\delta_a - \alpha_a)}{\alpha_i + p\delta})/\delta_u, (\delta_a - \alpha_a)/(\alpha_i + p\delta))^T$ are the left and right  eigenvectors for $\lambda_3=0$, respectively. Moreover,  ${\bf f}_\beta = (0,0, (1-a-u-i)i)^T$. A straightforward computation at the bifurcation point leads to

\vspace{0.25cm}

\noindent
1) ${\bf u} \cdot {\bf f}_{\beta}^{BP} = 0$, \\
2) ${\bf u} \cdot (D{\bf f}_{\beta}^{BP} {\bf v}) = v_3 = (\delta_a - \alpha_a)/(\alpha_i + p \delta) > 0$, and \\
3) ${\bf u} \cdot \left(D^2 {\bf f}^{BP}({\bf v}, {\bf v}) \right) = -2v_3 \left( (\beta-\beta_a) v_1 + (\beta-\beta_u) v_2 + \beta v_3 \right) < 0$.

\vspace{0.25cm}
\noindent
Note that the  inequality $\alpha_a < \delta_a$ and the assumption
$0 \leq \beta_a, \beta_u < \beta$ of the SAUIS model imply that all coordinates of~$\bf v$ and coefficients in~3) are positive, so that the whole expression becomes negative and, in particular, nonzero.
So, when $R^a_0 < 1$  system \eqref{eqn:SAUIS} experiences a transcritical bifurcation as $\beta$  crosses the bifurcation value $\beta=\delta$ \cite{Perko}.
Moreover,  Theorem 4.1 in~\cite{Castillo}, together with the inequality~$> 0$ in~2) and the inequality~$< 0$ in~3), implies that the direction of the bifurcation is always the same, namely, system~\eqref{SAUIS} experiences a forward bifurcation at~$R_0 = 1$.

\subsubsection{Transcritical bifurcation at $\lambda_3(P_2)=0$}

Let us now assume $\alpha_a > \delta_a$ to guarantee the existence of a positive $P_2$, and $\beta > \delta$ to allow $\lambda_3(P_2)$ to be positive for some parameters values. From the discussion surrounding~\eqref{eqn:P2lambda3+}  it follows that $\lambda_3(P_2)=0$ if and only if
\begin{equation}\label{BP}
\frac{\beta - \beta_a}{\beta - \delta} \, a^*_0 + \frac{\beta - \beta_u}{\beta - \delta} \, u^*_0 = 1,
\end{equation}
with $a^*_0$ and $u^*_0$ given by \eqref{eqn:P2-location}. That is, at this parameter combination, the endemic equilibrium $P_3$ bifurcates from $P_2$, the disease-free equilibrium with aware individuals.

When the Jacobian matrix~\eqref{JP2} is evaluated at the bifurcation point, the value in the lower right corner becomes zero. We get
\begin{equation*}
J^{BP}  =  \left( \begin{array}{ccc}
- \alpha_a\,  a^*_0 \ \, & \ \, -\alpha_a\, a^*_0 \ \, &  \ \, \alpha_i \delta_a/\alpha_a - (\alpha_a+\beta_a)a^*_0 + p \delta
\\
\delta_a + \alpha_u (\delta_a/\alpha_a - a^*_0)  \ \, & \ \, -\alpha_u a^*_0 - \delta_u  \ \, &  \ \, -\alpha_u a^*_0 - \beta_u u^*_0 + q \delta
\\
0 \ \,  & \ \, 0 \ \,  & \ \, 0
\end{array}
\right),
\end{equation*}
which has  the row vector ${\bf u} = (0,0,1)$  as left  eigenvector with eigenvalue $\lambda_3 = 0$.   Since the determinant $\det\left( J_2^{BP} \right) = \alpha_a a^*_0(\delta_a+\delta_u+\alpha_u\delta_a/\alpha_a) > 0$ of the submatrix $J_2^{BP}$ of $J^{BP}$ formed by the first two rows and the first two columns is strictly positive,  the first two columns of~$J_2^{BP}$ are linearly independent.  Therefore, the third component~$v_3$ of the right eigenvector  $\bf v$ with eigenvalue $\lambda_3=0$ must be different from $0$.

If we take $\mu=\beta_a$ as  the bifurcation parameter, then ${\bf f}_{\beta_a} = (-a\,i, 0, a\,i)^T$. As before, a straightforward computation at the bifurcation point leads to:

\vspace{0.25cm}
\noindent
1) ${\bf u} \cdot {\bf f}_{\beta_a}^{BP} = 0$, \\
2) ${\bf u} \cdot (D{\bf f}_{\beta_a}^{BP} {\bf v}) = a^*_0 v_3 \ne 0$, and \\
3) ${\bf u} \cdot \left(D^2 {\bf f}^{BP}({\bf v}, {\bf v}) \right) = -2v_3 \left( (\beta-\beta_a) v_1 + (\beta-\beta_u) v_2 + \beta v_3 \right)$.

\bigskip

\vspace{0.25cm}
\noindent
Recall that $\bf v$ is by definition orthogonal to the first two rows of~$J^{BP}$ and these are linearly independent. Thus $\bf v$ cannot also be orthogonal to any vector that is linearly independent of the latter two vectors. Therefore, since $v_3 \ne 0$, the inequality ${\bf u} \cdot \left(D^2 {\bf f}^{BP}({\bf v}, {\bf v}) \right) \ne 0$ will follow whenever the vector given by ${\bf b} := (\beta-\beta_a, \beta-\beta_u,  \beta )$ is linearly independent of the two first rows of~$J^{BP}$.  This may not always be the case, but it will be generically true.  For instance, the first two columns of~$J^{BP}$ are linearly independent and the parameter~$\alpha_i$ appears in only one position of~$J^{BP}$,  while $a_0^*$ and $u_0^*$ that give the location of~$P_2$ do not depend on it.  Since it  does not appear in~$\bf b$, one can take $\alpha_i$ as a free parameter in order to see whether the previous inequality fails for a particular positive choice of~$\alpha_i$ given any choice of all other parameters.

\begin{figure}
\begin{center}
\begin{tabular}{c}
\includegraphics[scale=0.45]{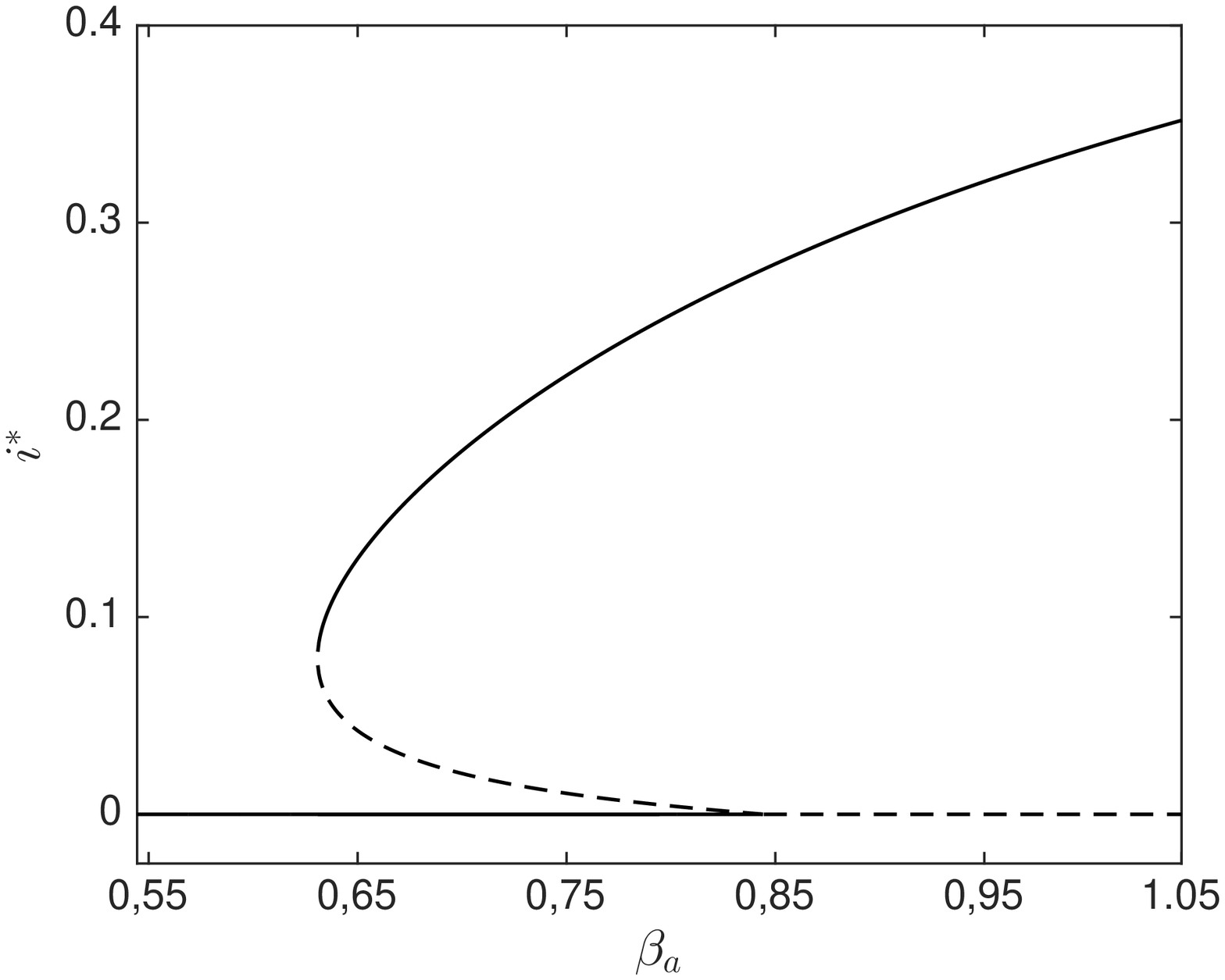}
\\
\includegraphics[scale=0.45]{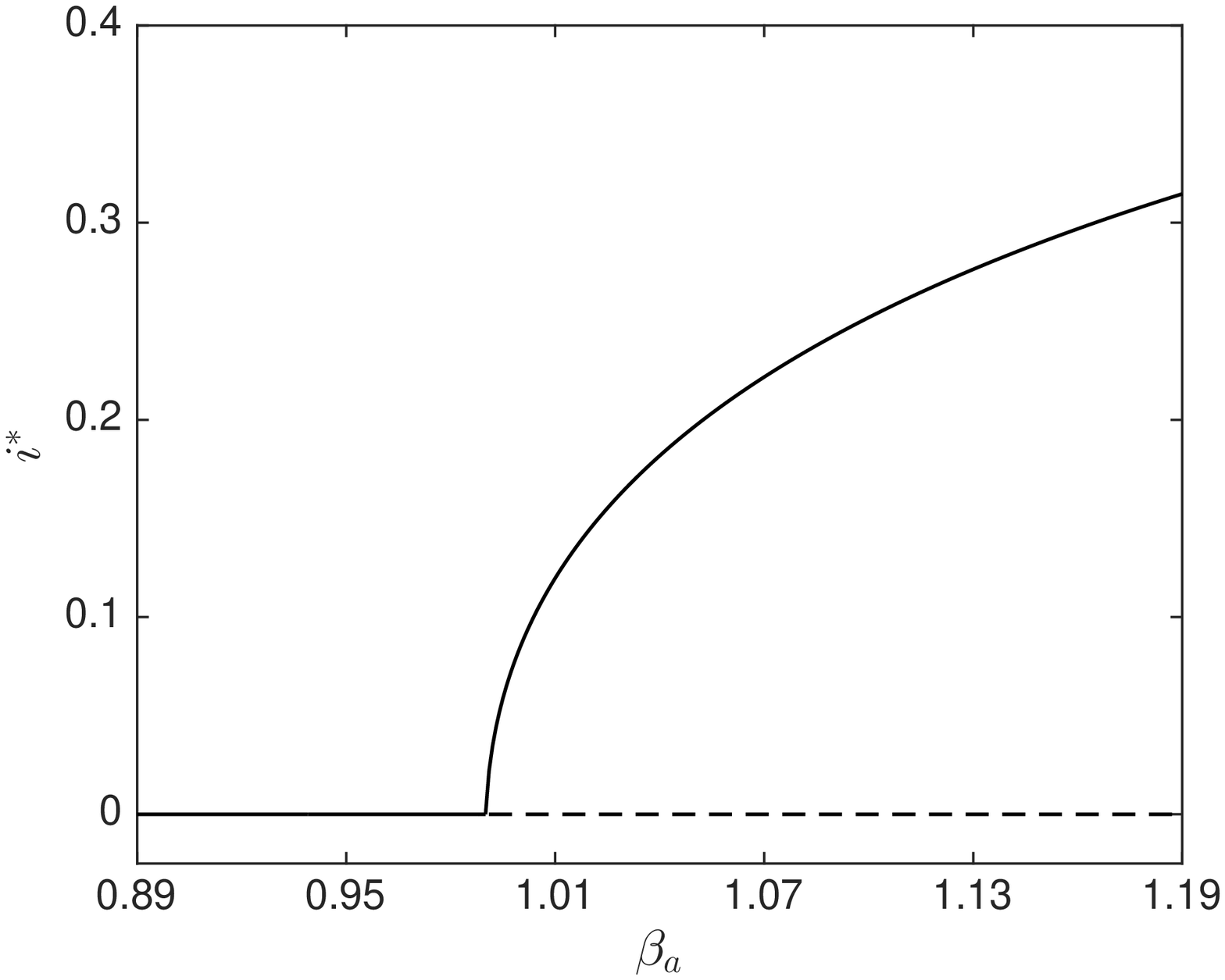}
\end{tabular}
\caption{Transcritical bifurcation diagrams of system \eqref{eqn:SAUIS} for $\beta=2$, $\beta_u=1$, $\delta=1$, $\delta_a=0.01$, $\delta_u=0.05$, $\alpha_i=0.8$, $\alpha_u=0.1$, $p = q = 0$, $\alpha_a=0.1$ (top) and $\alpha_a=1$ (bottom). The stable (unstable) equilibria are depicted with a solid (dashed) line. Bifurcation values: $\beta_a^c=0.844$ and $\beta=0.62$ (for the fold bifurcation) (upper panel); $\beta_a^c=0.988$ (lower panel).
\label{Fig:Transcritical}}
\end{center}
\end{figure}

Thus the criterion given in~\cite{Perko} after Sotomayor's Theorem guarantees that for a nonempty open set of parameter settings at which~$P_1 \neq P_2 \in \Omega$ the  system~\eqref{eqn:SAUIS} experiences a transcritical bifurcation as $\beta_a$ passes through the bifurcation value
$$
\beta_a^c :=\beta - \frac{1}{a_0^*} \left( \beta-\delta - (\beta-\beta_u)u_0^* \right).
$$
However, in contrast to what happens at $R_0=1$, the direction of the bifurcation is not always the same. To illustrate this fact, Figure~\ref{Fig:Transcritical} shows an example of forward and backward bifurcations occurring at $\lambda_3(P_2)=0$ for $\beta_a < \beta$.  Here $p = q = 0$, but similar examples with $p, q > 0$ exist.

The same conclusion holds if we use $\beta$ or $\beta_u$ as a  bifurcation parameter. However, the use of $\beta_a$ seems more appropriate because its value is related to the effectiveness of the response adopted by aware hosts.

\subsection{Hopf bifurcations}

We will see that in contrast to the SAIS model,  sustained oscillations are possible in the SAUIS model thanks to the occurrence of Hopf bifurcations.
A pair $(\sigma, \tau)$ of parameters of~\eqref{eqn:SAUIS} will be called a \emph{Hopf pair} if there exists an equilibrium point  at which the Jacobian matrix has a pair of pure imaginary eigenvalues. This definition of course depends on chosen values of the other parameters.  For simplicity we will always implicitly assume that these other parameters are fixed and suppress them in our notation.

An explicit criterion that specifies whether an $n \times n$ matrix~$M$, with coefficients that may depend upon parameters, has a pair of pure imaginary eigenvalues is given in \cite{GMS97}. For our purposes the case when $n = 3$ and $M=J$ is the Jacobian matrix at endemic equilibrium~$P_3$ is relevant. Let $p(\lambda)=\lambda^3 + c_2 \lambda^2 + c_1 \lambda + c_0$ be the characteristic polynomial of~$J$. Its coefficients are  $c_0 = -\det(J)$,  the sum ~$c_1$ of the principal minors of $J$, and $c_2 = - \text{trace}(J)$. According to Theorem 2.1 and Table 1 in~\cite{GMS97}, the matrix $J$ has precisely one pair of pure imaginary eigenvalues if and only if
\begin{equation} \label{Hopf}
   c_0 - c_1 c_2 = 0 \quad \text{and} \quad c_1 > 0.
\end{equation}

Therefore, to locate the Hopf pairs in a $(\sigma, \tau)$ parameter space, we will use~$\sigma$ as free parameter and solve the system given by  the equilibrium equations
{\small
\begin{equation}\label{equil}
\begin{split}
\alpha_i (1-a^*-u^*-i^*) i^* + \alpha_a (1-a^*-u^*-i^*) a^* - \beta_a a^* i^* - \delta_a a^* + p \delta i^* = 0 \\
\delta_ a a^* + \alpha_u (1-a^*-u^*-i^*) a^* - \beta_u u^* i^* - \delta_u u^*  + q \delta i^* = 0,
\end{split}
\end{equation} }
combined with~\eqref{i*} and~\eqref{Hopf},
for $a^*$, $u^*$, $i^*$, and $\tau$, while all other parameters are set to fixed values.

The component $\tau$ of the solution, if any, and the corresponding value of $\sigma$ define a Hopf pair $(\sigma, \tau)$ for which a Hopf bifurcation occurs at the endemic equilibrium $P_3=(a^*,u^*,i^*)$. The set of Hopf pairs defines the so-called Hopf-bifurcation curve $H$ in $(\sigma, \tau)$ parameter space,
$$
H = \{ (\sigma, \tau) \in \mathbb{R}^2_+ \, | \, \exists P_3 \in \mathbb{R}^3_+ \, \text{for which} \, \eqref{i*}, \, \eqref{Hopf}, \, \text{and} \, \eqref{equil} \, \text{hold} \}.
$$
Note that $H$ can be parametrized by $i^*$ (or any of the components of $P_3$) \cite{SSK12}.

\begin{figure}
\begin{center}
\includegraphics[scale=0.5]{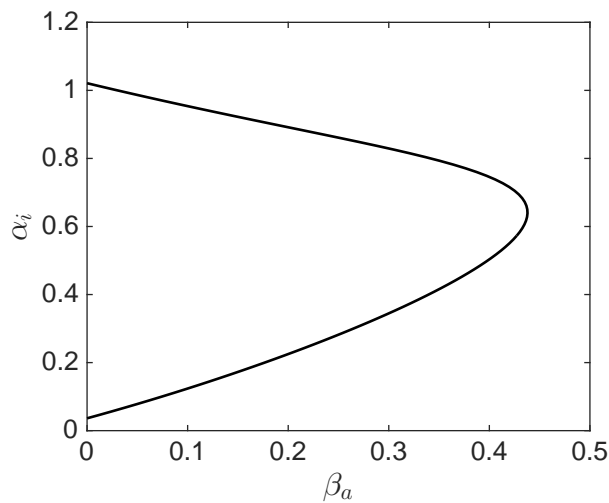}
\caption{Hopf-bifurcation curve $H$ of system \eqref{eqn:SAUIS} for $\delta=1$, $\delta_a=0.01$, $\delta_u=0.05$, $\beta=3$, $\beta_u=0.5$, $\alpha_a=0.01$, $\alpha_u=1$, and $p = q = 0$. For pairs $(\beta_a,\alpha_i)$ inside the region bounded by this curve and the $\alpha_i$-axis system \eqref{eqn:SAUIS} has an unstable endemic equilibrium and a stable periodic orbit.}
\label{Fig:Hopf-values-aiba}
\end{center}
\end{figure}

\begin{figure}
\begin{center}
\includegraphics[scale=0.5]{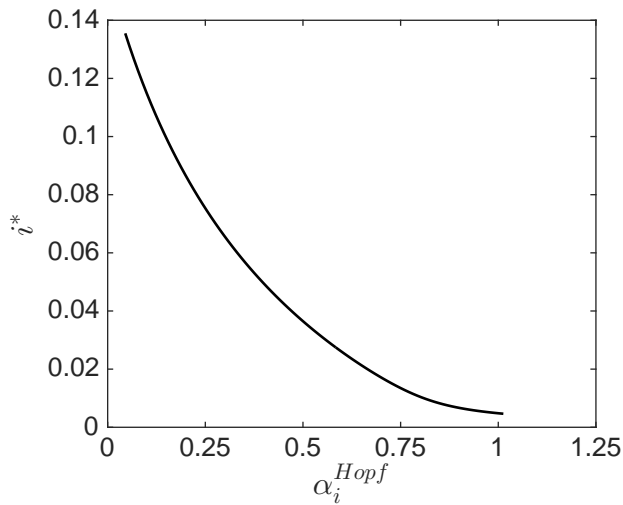}
\caption{Fraction of infectious hosts as a function of $\alpha_i$ along the Hopf-bifurcation curve $H$ of system \eqref{eqn:SAUIS} for $\delta=1$, $\delta_a=0.01$, $\delta_u=0.05$, $\beta=3$, $\beta_u=0.5$, $\alpha_a=0.01$, $\alpha_u=1$, and $p = q = 0$. At both ends of the curve, $\beta_a=0$.}
\label{Fig:Hopf-infectious-ai}
\end{center}
\end{figure}

We will focus here on the case $\sigma = \beta_a$ and $\tau = \alpha_i$. This choice appears to be of particular interest, as~$\alpha_i$ may be most amenable to  alteration by increased epidemiological monitoring (see Section~\ref{sec:Discussion}), while $\beta_a$ can be thought of as inversely proportional to the effectiveness of the behavioral response in aware hosts (see Section~\ref{sec:Discussion}). An example of a curve $H$ in the $(\beta_a, \alpha_i)$ parameter space is shown in Figure~\ref{Fig:Hopf-values-aiba}. For each value of $\beta_a$ within the range $[0, 0.4379)$, there exist two solutions of equations~\eqref{Hopf} and~\eqref{equil}.
The prevalence of the disease at the bifurcation points along the curve $H$ is presented in Figure~\ref{Fig:Hopf-infectious-ai}. The figure clearly shows how $i^*$ decreases monotonously with the value of $\alpha_i$ at the Hopf pairs.

\begin{figure}
\begin{center}
\begin{tabular}{cc}
\hspace{-1cm}
\includegraphics[scale=0.35]{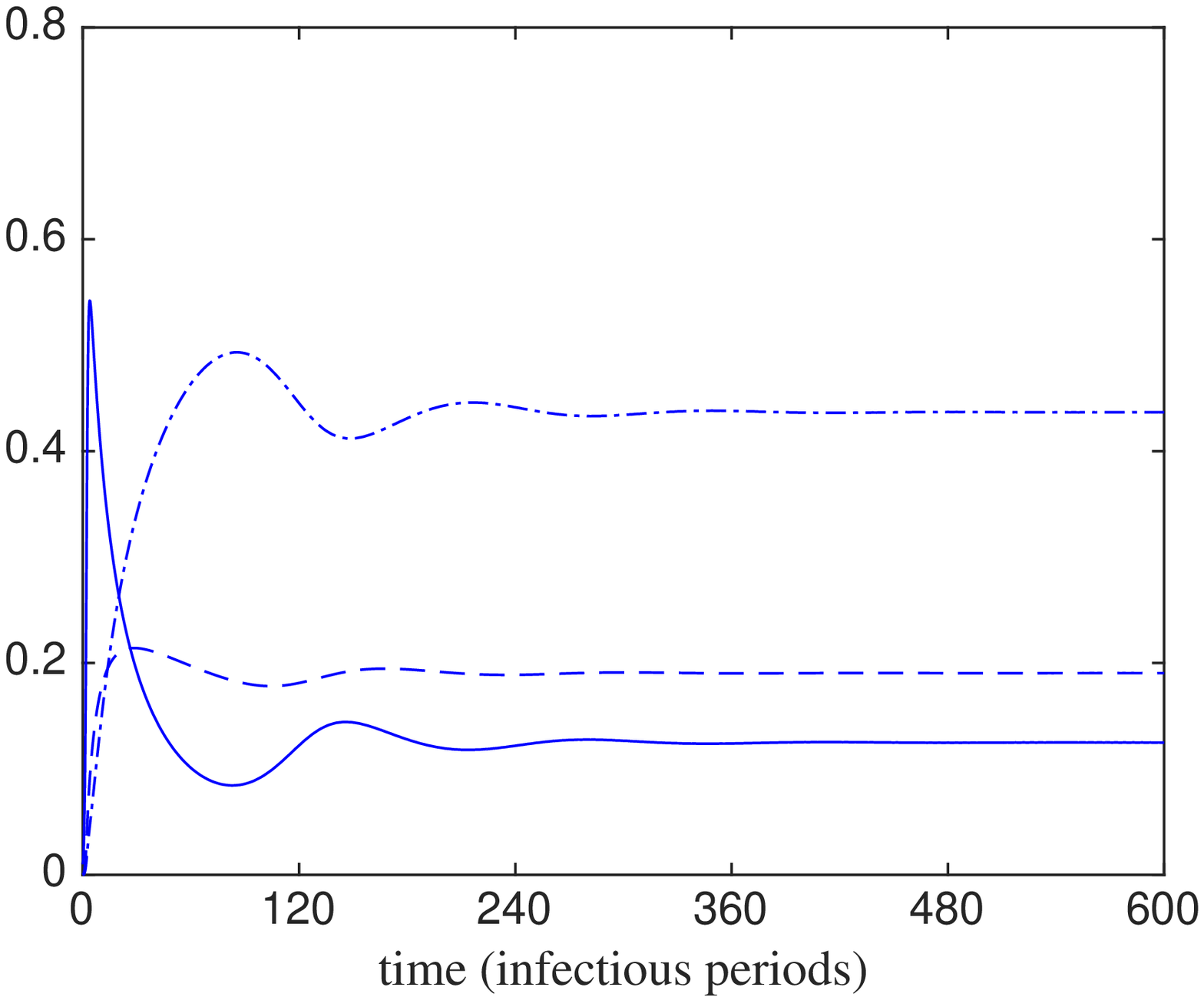}
&
\hspace{-0.75cm}
\includegraphics[scale=0.35]{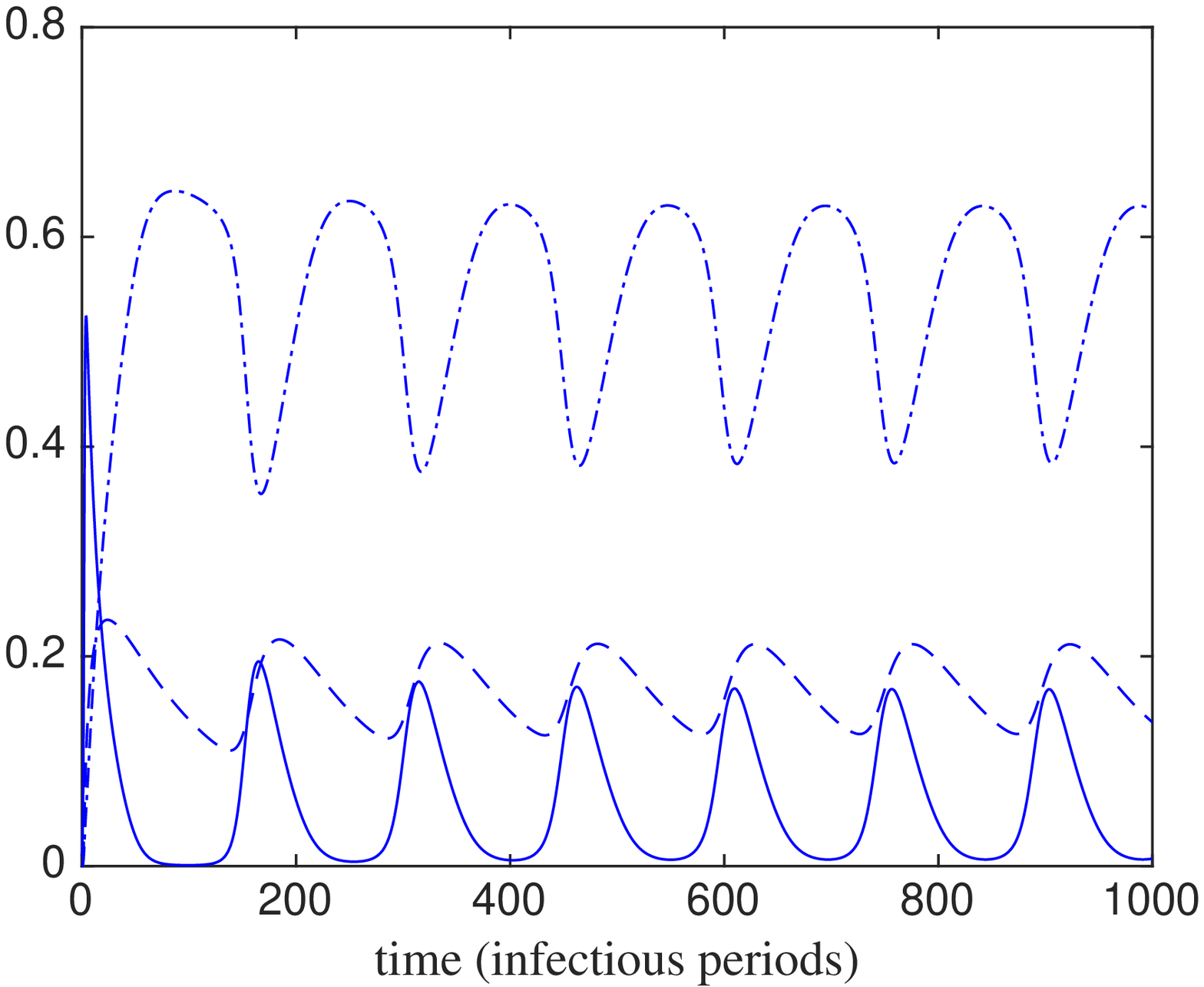}
\\
\hspace{-1cm}
\includegraphics[scale=0.35]{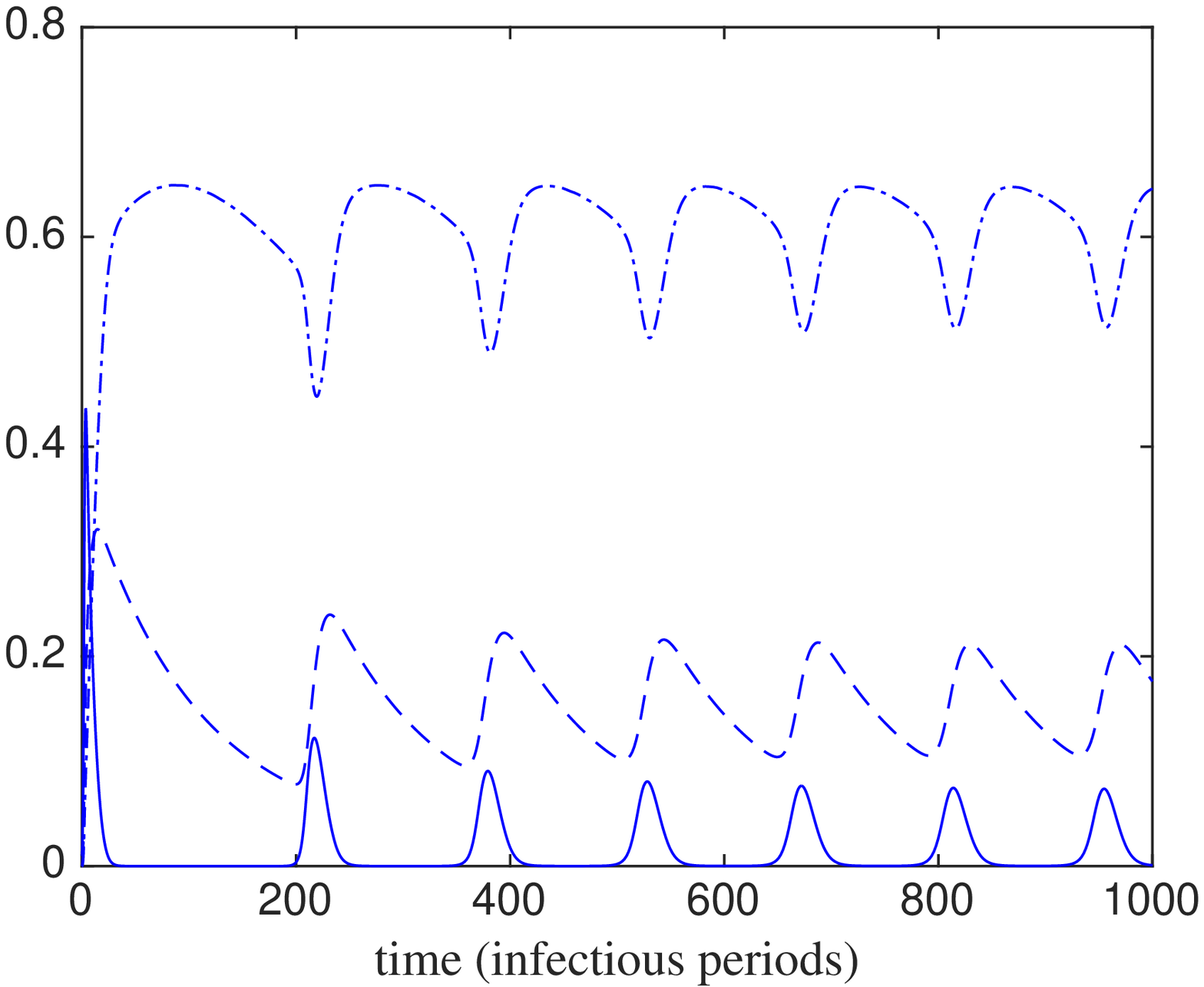}
&
\hspace{-0.75cm}
\includegraphics[scale=0.35]{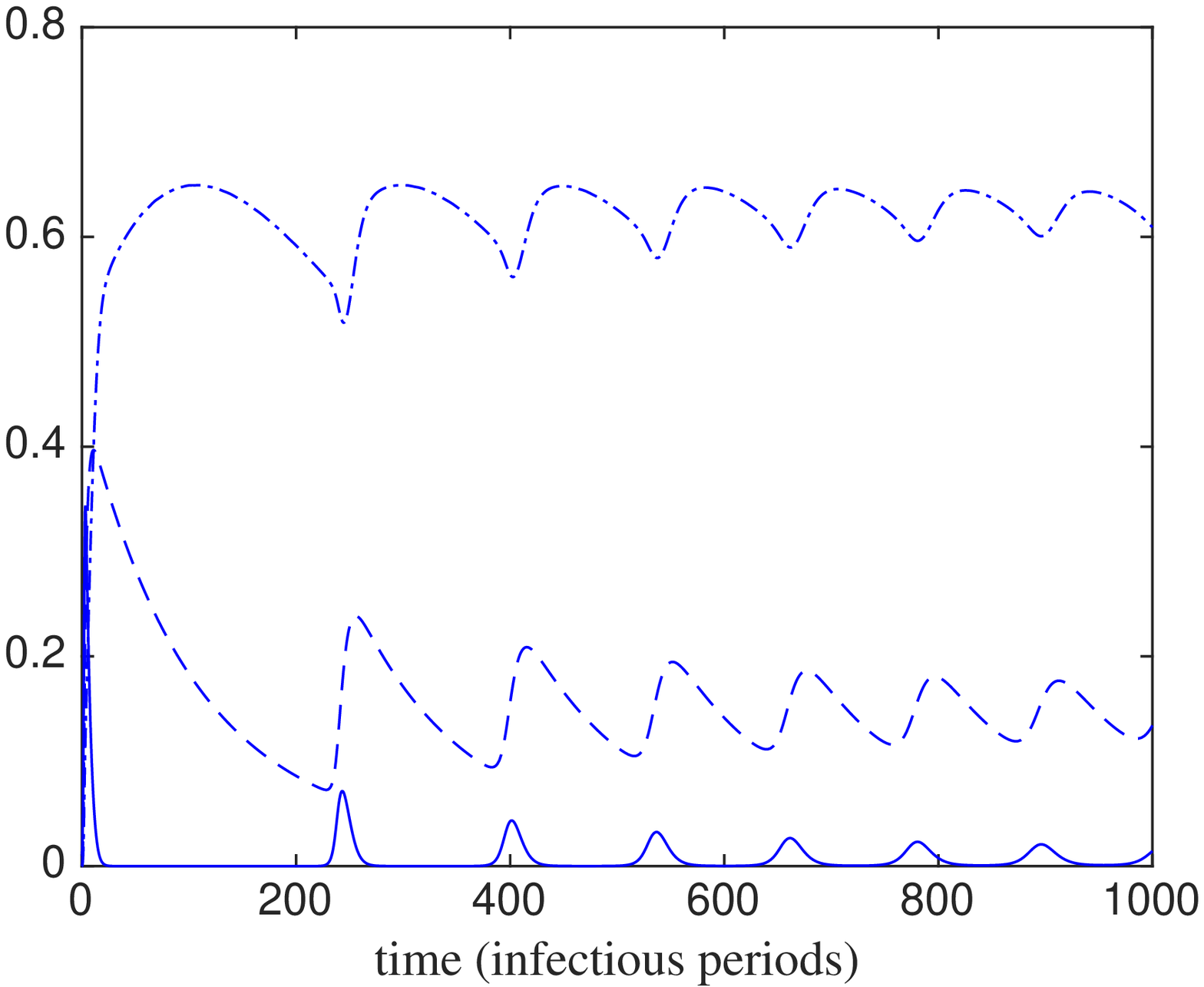}
\end{tabular}
\caption{Evolution of the fraction of infectious (solid line), aware (dashed line), and unwilling (dot-dashed line) hosts according to system \eqref{eqn:SAUIS} for different values of $\alpha_i$ along the vertical section in Figure~\ref{Fig:Hopf-values-aiba} corresponding to $\beta_a=0.2$: $\alpha_i=0.2$ (top left), 0.24 (top right), 0.5 (bottom left), 0.94 (bottom right). Fixed parameters: $\alpha_a = 0.01, \ \alpha_u = 1, \ \delta = 1, \ \delta_a = 0.01, \ \delta_u = 0.05, \ \beta = 3, \ \beta_a = 0.2, \ \beta_u = 0.5, \ p = q = 0$. Initial condition:  $a(0) = u(0) = 0$, $i(0) = 0.1$.
\label{Fig:Evolution-SAUIS}}
\end{center}
\end{figure}

The (sub- or supercritical) character of the Hopf bifurcation can be found by computing the first Lyapunov exponent, which is close to $-2.5$ for intersection of the straight line $\beta_a=0.2$ with the upper branch of the curve of predicted bifurcation points and close to~$-0.25$ for the intersection of this line with the lower branch.
Thus for each pair $(\beta_a,\alpha_i)$ inside $\mathcal{R}$, the endemic equilibrium $P_3$ is unstable, as $J(P_3)$ has two complex eigenvalues with positive real part and a third eigenvalue that is real and negative, while stable periodic orbits appear around $P_3$. Outside this region, $P_3$ is asymptotically stable. Therefore, the points on the curve $H$ define supercritical Hopf bifurcations of the system~\eqref{eqn:SAUIS}. The numerical explorations in Figure~\ref{Fig:Evolution-SAUIS} confirm this prediction.

Numerical simulations confirm our predictions regarding the Hopf pairs shown in Figure~\ref{Fig:Hopf-values-aiba}. In particular, for $\beta_a = 0.2$ and the parameter settings as specified in the caption of Figure~\ref{Fig:Hopf-values-aiba}, Hopf bifurcations in system \eqref{eqn:SAUIS} are predicted for $\alpha_i^* = 0.2251$ and $\alpha_i^{**} =  0.8916$.  Figure~\ref{Fig:Evolution-SAUIS} illustrates what happens if we increase $\alpha_i$ from below~$\alpha_i^*$ to above~$\alpha_i^{**}$ for trajectories with initial condition $a(0)=u(0)=0$ and $i(0)=0.1$,  far from the endemic equilibrium~$P_3$.

The top left panel of Figure~\ref{Fig:Evolution-SAUIS} shows a simulation with $\alpha_i = 0.2$, right below the first Hopf bifurcation. The solution quickly approaches an endemic equilibrium with disease prevalence $i^* =  0.1249$. When we increase $\alpha_i$ slightly above~$\alpha_i^*$, to~$0.24$, then we observe significant oscillations in the top right panel. Moreover, these oscillations correspond to a stable limit cycle in the phase portrait of the system that attracts other trajectories whose initial conditions are not necessarily close to the endemic equilibrium. For the (unstable) endemic equilibrium we get  $i^* = 0.0651$ in this case. Due to the oscillations, the long-term mean prevalence will be even slightly lower, around~$0.0592$.

A similar picture of persistent oscillations is shown in the bottom left panel of Figure~\ref{Fig:Evolution-SAUIS} where we chose $\alpha_i = 0.5$, about half-way between~$\alpha_i^*$ and~$\alpha_i^{**}$.  The existence of a stable limit cycle becomes clearly noticeable in simulations over a longer time horizon. Here the prevalence at endemic equilibrium is~$i^* = 0.0148$, very close to the mean prevalence $\overline{i} = 0.0150$ in the long run.  Note that for these settings long periods of very low disease prevalence alternate with short  spikes that indicate periodic small flare-ups.  These flare-ups are followed by a rapid increases in awareness, which indicate a panic-like spread of information.

Finally, when we further increase~$\alpha_i$ beyond~$\alpha_i^{**}$ to~0.94, we initially observe a similar pattern of alternating periods of extremely low disease prevalence, interrupted by small flare-ups with panic-like spread of awareness (see the bottom right panel in Figure~\ref{Fig:Evolution-SAUIS}).  The prevalence at endemic equilibrium further decreases to~$i^* = 0.0065$.  However, as this panel indicates, the amplitude of these oscillations now decreases, albeit very slowly, because the pair $(\beta_a, \alpha_i)=(0.2, 0.94)$ is very close to the curve $H$.

\begin{figure}
\begin{center}
\begin{tabular}{cc}
\hspace{-0.5cm}
\includegraphics[scale=0.34]{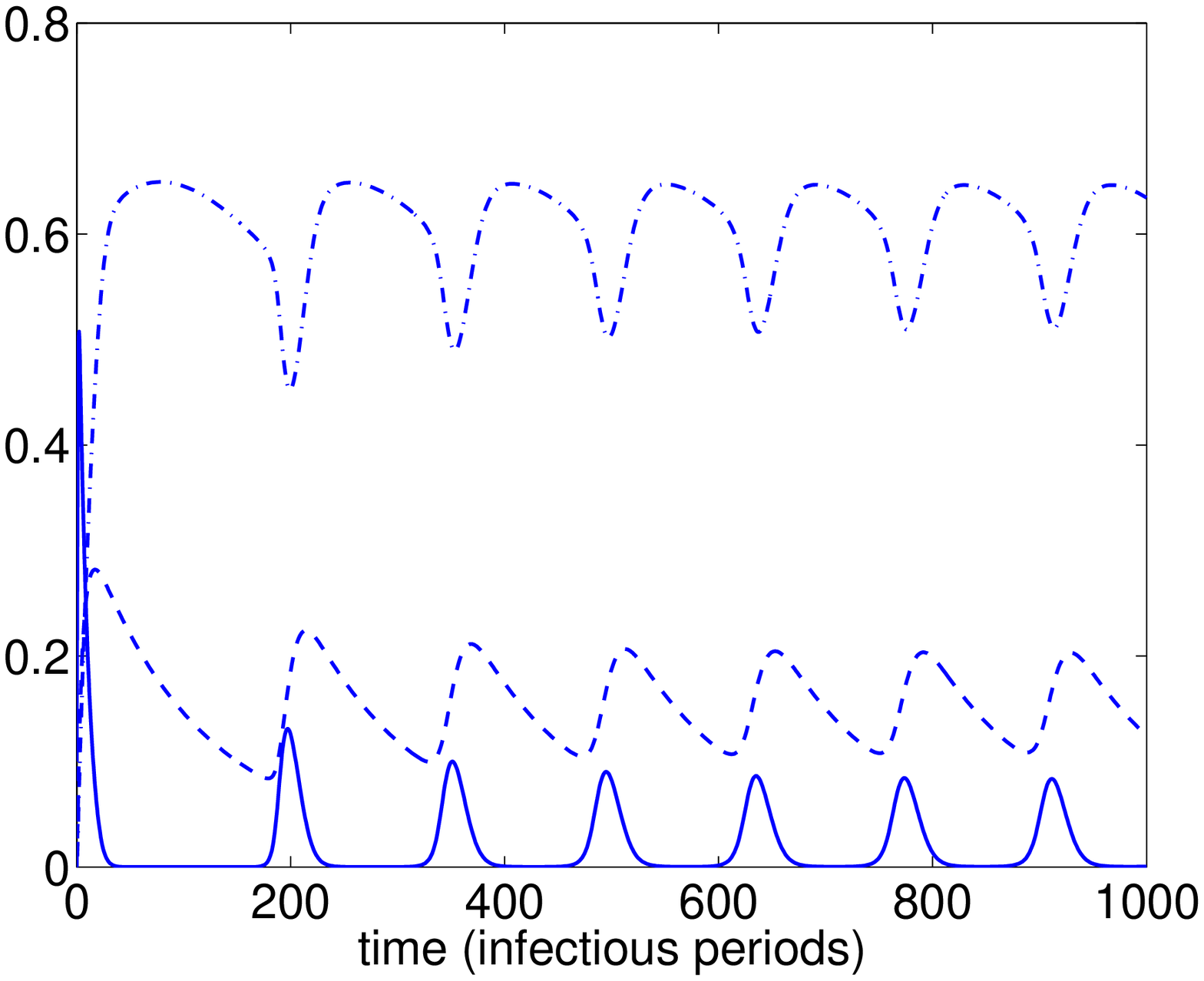}
&
\hspace{0.2cm}
\includegraphics[scale=0.34]{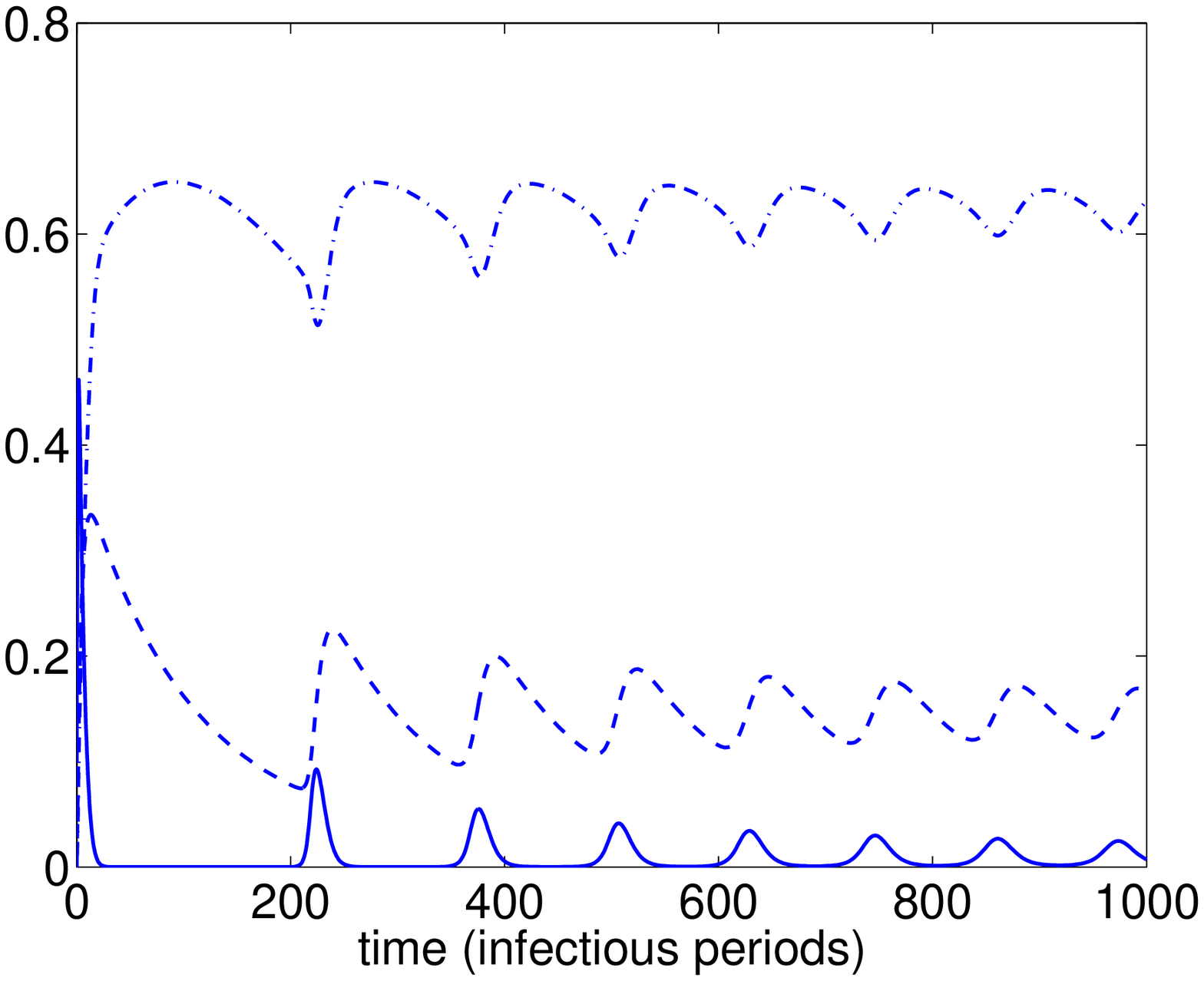}
\\
\hspace{-0.5cm}
\includegraphics[scale=0.34]{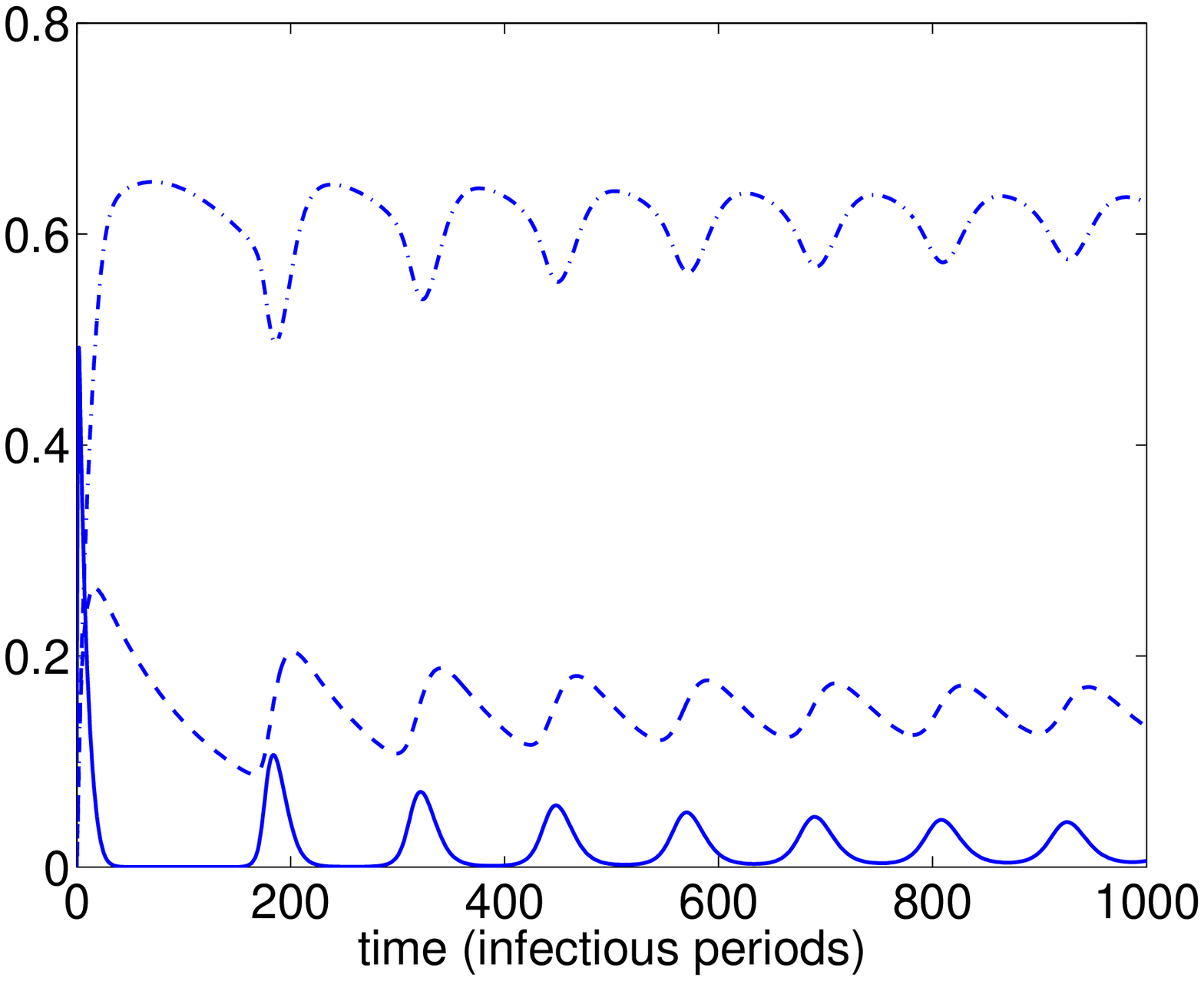}
&
\hspace{0.2cm}
\includegraphics[scale=0.34]{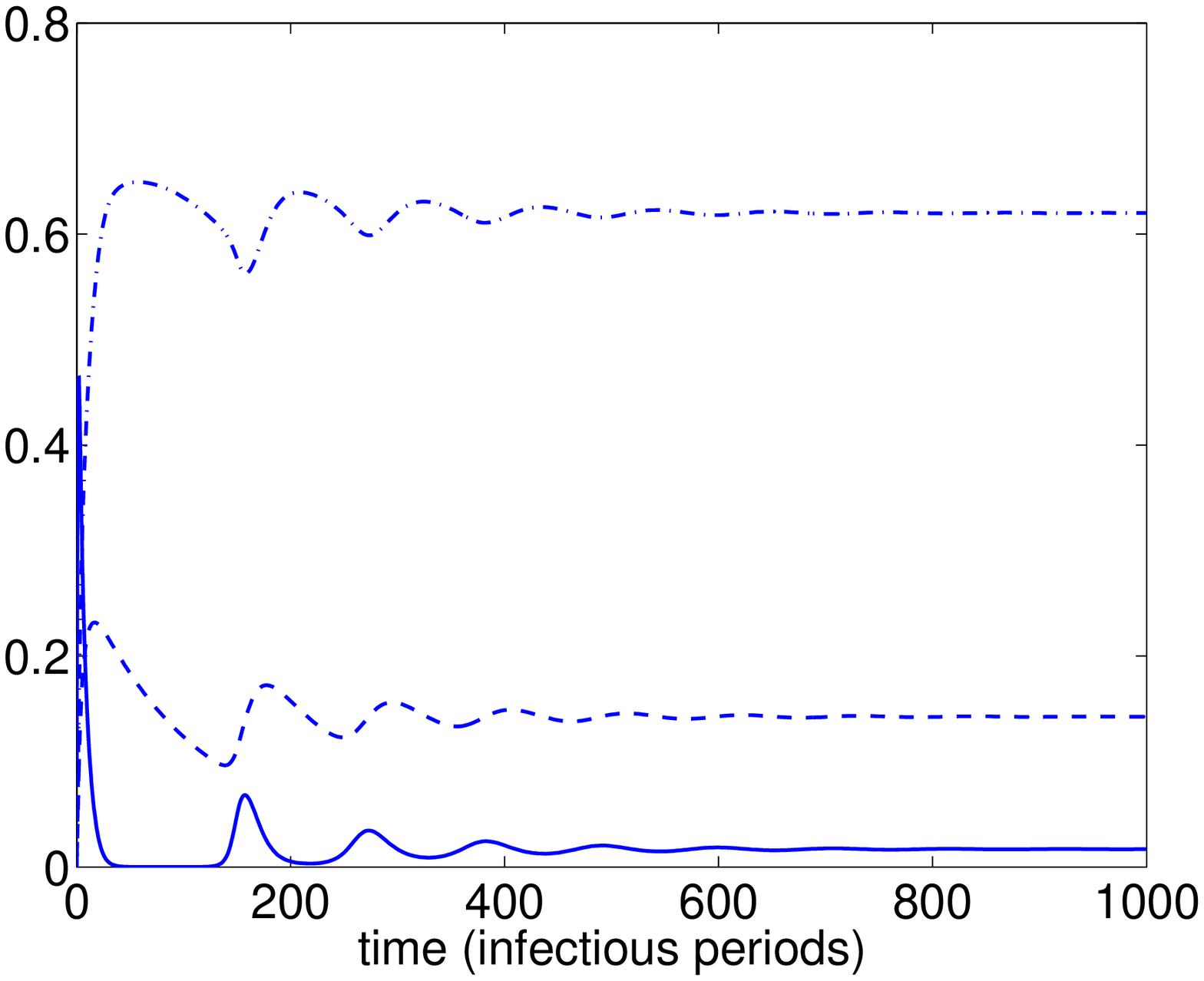}
\end{tabular}
\caption{Evolution of the fraction of infectious (solid line), aware (dashed line), and unwilling (dot-dashed line) hosts for different values of $q$ corresponding to $p=0.05$; $q=0.05$ (top left), 0.1 (bottom left), 0.2 (bottom right).  Initial condition:  $a(0) = u(0) = 0$, $i(0) = 0.1$.  In the left and bottom right panels, $\alpha_u = 1$, $\delta = 1$, $\delta_a = 0.01$, $\delta_u = 0.05$, $\beta = 3$, $\beta_u = 0.5$, $\alpha_a = 0.01$, $\beta_a=0.2$ and $\alpha_i=0.2$.  In the top right panel, $\alpha_i = 0.24$, $p = 0.1 $, $q = 0.1$, and all the other parameters stay the same.
\label{Fig:Evolution-p-q}}
\end{center}
\end{figure}

It is interesting to compare Figure~\ref{Fig:Evolution-SAUIS} with  Figure~\ref{Fig:Evolution-p-q}.  The parameter settings in the latter are similar to those in the upper left panel of Figure~\ref{Fig:Evolution-SAUIS}, except that now $p$ and $q$ assume small positive values. In particular, compare the upper left panels of these figures. While sustained oscillations are absent when $p = q = 0$ (Figure~\ref{Fig:Evolution-SAUIS}), they do occur when $p = 0.05$ and $q$ is very small. As the two lower panels of Figure~\ref{Fig:Evolution-p-q} show, increasing~$q$ first dampens and then eliminates these oscillations.  Thus we conclude that for these settings of the remaining parameters the oscillations are driven by having a positive proportion of hosts who move into the A-compartment upon recovery. The upper right panel of Figure~\ref{Fig:Evolution-p-q} corresponds to the upper right panel of Figure~\ref{Fig:Evolution-SAUIS} and shows similar oscillations, but with decreased amplitude when $p, q > 0$.

\begin{figure}
\begin{center}
\begin{tabular}{c}
\includegraphics[scale=0.42]{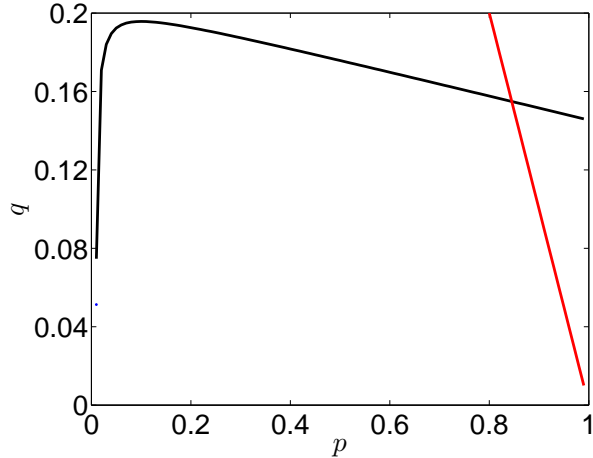}
\end{tabular}
\caption{Hopf-bifurcation curve H of system \eqref{eqn:SAUIS}. The straight line sets the boundary $p + q = 1$, so only the part of the Hopf-bifurcation curve that lies on the left of this line makes biological sense.  When $p = 0.05$, $q^*=0.1923$ is the only Hopf-bifurcation point.  Here, $\alpha_u = 3$, $\delta = 1$, $\delta_a = 0.01$, $\delta_u = 0.05$, $\beta = 3$, $\beta_u = 0.5$, $\alpha_a = 0.012$, $\beta_a = 0.2$, $\alpha_i = 0.05$.
}
\label{Fig:p-q-Hopf}
\end{center}
\end{figure}

While~$p$ and~$q$ are fairly small in the parameter settings for Figure~\ref{Fig:Evolution-p-q}, we found that sustained oscillations are possible even when $p+q = 1$. Figure~\ref{Fig:p-q-Hopf} shows the Hopf bifurcation curve for the Hopf pair $(p,q)$ for a similar setting of the other parameters. The first Lyapunov exponent for the point on the Hopf bifurcation curve with $p = 0.1$ is close to $-2.6$.  Thus sustained oscillations occur in the area under the curve. The line segment inside this region indicates locations where $p+ q = 1$. Interestingly enough, for $q$ around $0.18$, the curve predicts no oscillations for values of~$p$ that are very small or very close to~1, while it predicts sustained oscillation when $p$ takes moderate values.
For more explorations of the influence of parameters~$p$ and~$q$ on the dynamics of the model, see~\cite{Xin}.

\section{Discussion}\label{sec:Discussion}

Previous results on various types of SAIS models had suggested that a behavioral response of hosts who are aware of an ongoing outbreak of a disease can be effective in eliminating the endemic equilibrium or driving it to very low levels. For \textit{non-reactive} SAIS models ($\alpha_a = 0$) it was shown in \cite{Sahneh12a} that if there is no awareness decay $(\delta_a = 0)$, then there is an elevated epidemic threshold  so that all outbreaks will be minor as long as $\beta_a < \delta < \beta$ and $\alpha_i$ is sufficiently large.
In \cite{Juher14}, it was proved that this elevation of the epidemic threshold disappears if one assumes that awareness will decay over time.  The reason is that in order to suppress epidemic spreading, we need a permanent fraction of aware hosts.
If awareness is eventually lost, then the presence of aware individuals created at the early stages of an epidemic is not enough to contain an outbreak, and the classic disease-invasion epidemic threshold $R_0 = \beta/\delta = 1$ drives the dynamics.

Here we have introduced and studied \emph{reactive} SAIS models that make the natural assumption that awareness can also be transmitted from an aware host to a susceptible one ($\alpha_a > 0$). We have shown that under certain conditions in these models  a permanent fraction of aware individuals can be sustained even in the presence of  awareness decay, which again can lead to elevation of the epidemic threshold above~$R_0$.

The question of when the response will be sufficient to prevent future flare-ups from low endemic levels  had not previously been addressed in the literature. This question is of interest for models  where awareness will decay over time. Lemma~\ref{NCO} of Section~\ref{SAIS-Non}  indicates that when there is not much differentiation between hosts in their propensity to share information about the disease with other hosts, future flare-ups are ruled out. This result applies both to SAIS models  with constant rates and to SAIS models in which the rate coefficients depend on the prevalence of the disease.

However, in most real human populations hosts will significantly differ in how effectively they contribute to growing awareness of other hosts.  We constructed a new class of models, SAUIS models, that incorporate this phenomenon.  We think of them as the simplest possible straightforward generalization of SAIS models in this direction.    SAUIS models permit persistent  cycles of flare-ups  that induce panic-like spread of information, which will temporarily drive the prevalence to low levels without leading to elimination of the disease or providing permanent protection against future major outbreaks. What we observe in such dynamics are flare-ups of the disease, closely followed by steep rises of hosts who are willing to spread awareness, which in turn are followed by increases in the number of unwilling hosts and corresponding decreases in the size of the S-compartment. This drives the disease prevalence to low levels while the numbers of unwilling hosts stay high, but the spread of awareness slows down until the S-compartment gets sufficiently replenished for the next flare-up of the disease.

This phenomenon does not depend on variable rate coefficients, as all of these are assumed constant in the version of SAUIS models presented here. It can occur regardless of whether some or all hosts are assumed to move into the union of the A-compartment and the U-compartment at the time of recovery from the disease.
While increasing the fraction of hosts who move to the U-compartment tends to dampen and eventually prevent oscillations, increasing the fraction of hosts who move into the A-compartment as a result of direct experience can sometimes induce and sometimes prevent oscillations.  This, together with the contrasting result for the closely related SAIS models, clearly demonstrates that the observed oscillations are in fact driven by the unequal propensity of hosts to share information  when differences in the willingness to share information arise from the degradation of the information during subsequent transmission events as modeled in~\cite{Funk09}.

These findings seem obviously relevant from the point of view of designing effective policies for controlling infections.  As the two top panels of Figure~\ref{Fig:Evolution-SAUIS} suggest and our calculations confirmed, the average disease load, as interpreted as the mean prevalence calculated from the area under the solid curve, may in the long run be  similar and sometimes even lower in the presence of periodic flare-ups than for similar parameter settings where an endemic equilibrium is approached. Thus if the primary goal of control is a reduction of average load, then elimination of periodic flare-ups may sometimes even be counterproductive.  However, peak load may be more important than average load in terms of the danger of overwhelming the health care  system~\cite{Nina},  and it will be higher under oscillatory dynamics. Thus if the primary goal of control is to reduce peak load of the disease as much as possible, then control measures should be targeted at elimination of possible flare-ups.

One possible control strategy is to commit resources to continuous monitoring the prevalence~$i$  of the disease and dissemination of this information by health care professionals and the media.  In terms of SAUIS models, this can be interpreted as increasing the parameter~$\alpha_i$. It seems plausible that this strategy will decrease the value of~$i^*$ at the endemic equilibrium and the long-term average disease load~$\bar{i}$. Our numerical explorations suggest that this intuition is correct.  When $\alpha_i$ crosses the threshold represented by the lower branch of the curve in Figure~\ref{Fig:Hopf-values-aiba}  the endemic equilibrium will become unstable, with oscillatory dynamics for values above the threshold. However, as Figure~\ref{Fig:Evolution-SAUIS} shows, right above the threshold the resulting oscillations may result in an increased peak load relative to values of~$\alpha_i$ right below it.  A further increase of $\alpha_i$ beyond a second threshold represented by the upper branch of the curve in Figure~\ref{Fig:Hopf-values-aiba} will lead to renewed local asymptotic stability of the endemic equilibrium.  But as the lower-right panel Figure~\ref{Fig:Evolution-SAUIS} shows, even above this threshold damped oscillations may remain observable for a considerable time horizon when  trajectories start far away from the equilibrium.
These and other results presented in Section~\ref{SAUIS} reveal a surprisingly rich and intricate dynamics of SAUIS models.  It becomes clear that optimizing the dynamics by controlling one or more parameters that govern the spread of awareness will usually be quite delicate even if one could treat the model at face value.

Our findings open several avenues of future research.  It would be interesting to analytically derive necessary and/or sufficient conditions on the information flow about the disease between different types of hosts that would preclude the existence of sustained oscillations.  However, in order to base actual public health policy on such results, they would need to be sufficiently robust and carry over to wider classes of models of transmission of awareness and degradation of information. The SAUIS models that we introduced and explored here are useful for clearly demonstrating effects that are precluded by the simplifying assumptions of previously studied SAIS models. However, in biologically more realistic models one would presumably want to consider more fine-grained scenarios of degradation of information, perhaps by incorporating a chain of compartments $A_1, A_2, \dots , A_m$ that would represent progressively decreasing willingness to share the information and/or decreasing strength of the behavioral response.  Such chains could also be used to more directly adapt the model of~\cite{Funk09}.

Moreover, the assumption of constant rate coefficients is clearly an oversimplification.  We already discussed in the context of SAIS models how and why $\alpha(i), \alpha_a(i), \delta_a(i)$ might depend on the actual prevalence~$i$. While at low levels of prevalence the  information received by susceptible hosts is unlikely to be confirmed by first-hand knowledge of actual cases, at  moderate or high prevalence such confirmation by direct observation is more likely. Thus if such confirmation determines whether receiving information moves a given host into the A-compartment or the U-compartment, the ratio of newly created aware to newly created unwilling hosts may increase with the prevalence of the disease.

It is also of interest to investigate to what extent differences in the propensity to share information play a role in either inducing or enhancing sustained oscillations when the underlying disease dynamics is, for example, of type SIRS rather than SIS, or if demographics are included in the model.

Another possible direction of future research is to recast our models in the framework of stochastic processes.  This might allow us to address the question of how long it takes to reach the absorbing disease-free state when populations are relatively small and trajectories are predicted to visit states with very low disease prevalence. It also would allow us to study the spread of the disease and of awareness on the relevant contact networks, similar to the work on nonreactive SAIS models
in~\cite{Sahneh14}.

Predictions of oscillatory dynamics  have been reported for a number of previously studied models of behavioral epidemiology,  both for models that incorporate demographics, and for models that, similarly to the ones studied here, ignore demographics.  However, to the best of our knowledge, variability in the propensity of hosts to further disseminate awareness  had not yet been identified as a possible driving mechanism for this phenomenon.  We conclude this paper with a brief discussion of some related previous results. This review is not intended to be exhaustive; our goal is only to illustrate the variety of other mechanisms that appear to be capable of  generating oscillations in disease prevalence.

Damped oscillations can be observed in non-reactive~SAIS models when an asymptotically stable interior equilibrium has complex eigenvalues~\cite{Juher14}.
They are also possible when the underlying disease dynamics is of type~SIR without demographics.
For example~\cite{EPCH}, considers spatially structured models where flight from endemic regions is identified as a key factor that will drive the overall dynamics.

Hopf bifurcations and sustained oscillations have been found in SEI- and SIR-based models with demographics~\cite{SEI} or with the related mechanism of prevalence-dependent recruitment of susceptibles into a core group~\cite{Velasco}. It is interesting to note that while the model of~\cite{SEI} incorporates a reaction to media coverage, a type of awareness, it admits Hopf bifurcations even when the parameter that represents the strength of this reaction is zero. Sustained oscillations can also occur in models with demographics where the behavioral response represents compliance with vaccination and the driving mechanism involves real or perceived benefits of failing to adopt the behavioral response~\cite{Bauch,dOno1,dOno2,Reluga}. Such benefits are absent in SAUIS models.  They were also reported for models of vaccination compliance for influenza with fixed populations~\cite{Brebanal,Vardavas}.  In these models, vaccination decisions are made from year to year for different strains, so that the underlying disease dynamics is closer to type SIS than SIR. There is no direct information transfer in these models, and decisions are based on experience with past outbreaks rather than on current incidence levels. The latter features are also present in the models of~\cite{dOno1,dOno2}.

Sustained oscillations also have been reported for a model that in some aspects resembles a reactive SAIS model, but makes an assumption about a fixed number of susceptibles.  The authors of that paper conjectured that this assumption is needed for sustained oscillations in their model (see page~1353 of~\cite{Twitter}).

In~\cite{Grassly} it is argued that oscillations observed in longitudinal studies of incidence data of syphilis may be entirely explainable by the underlying disease dynamics of type SIRS even though clear patterns of behavioral changes over time were present.  This explanation is not consistent with an underlying disease dynamics of type SIS, and the authors of~\cite{Grassly} partly base their conclusion on the absence of such oscillations in corresponding data sets on gonorrhea for the time frame of 1941--2001. However, as we mentioned in the introduction, a recent increase in the incidence of gonorrhea and other STIs that appears to be driven by changing behavior patterns has been reported~\cite{Wilton}.

Finally, there exists a substantial literature on oscillations in models with a behavioral response that involve rewiring the contact network (see, for example,
\cite{Gross06,RZ,ShSch,Zhou}).  While these are individual-based models, analytical confirmation can sometimes be obtained by coarse-graining~\cite{GrK} or using pair-approximations to build corresponding ODE-models with nonuniform mixing~\cite{Szabo,SSK12}. Thus the mechanism that drives oscillations in these models is very different from the one in our SAUIS models, where uniform mixing is implicitly assumed.

\section*{Acknowledgments}
The work of J.S. has been partially supported by the research grants MTM2014-52402-C3-3-P of the Spanish government (MINECO) and MPCUdG2016/047 of the Universitat de Girona. J.S. is member of the research group 2014 SGR 1083 of the Generalitat de Catalunya.  We thank both referees for valuable suggestions that helped us in greatly improving this paper.

\section*{Appendix: Proof of Lemma~\ref{lem:unique}}

\noindent \textit{Proof of Lemma~\ref{lem:unique}.}
We can rewrite the equation~\eqref{eqn:i-null-(a)} of the $i$-nullcline as
\begin{equation}\label{eqn:i-nullcline-inv}
a_1(i) = \frac{\beta - \delta}{\beta- \beta_a} -\frac{\beta}{\beta - \beta_a}i.
\end{equation}

We will be using subscripts~1 and~2 in this proof to distinguish the function defined by~\eqref{eqn:i-nullcline-inv} from the right-hand side
of \eqref{eqn:a-null-pos} that defines the positive branch~$a_2(i)$ of the $a$-nullcline:
{\small
\begin{eqnarray}\label{eqn:a-null-pos2}
a_2(i) & = & \frac{1}{2} \, \left( 1 - i - \frac{(\alpha_i(i) + \beta_a) i + \delta_a(i)}{\alpha_a(i)} \right. \\
& & \left.
+ \   \sqrt{\left(  1 - i - \frac{(\alpha_i(i) + \beta_a) i + \delta_a(i)}{\alpha_a(i)} \right)^2 + 4 \frac{\alpha_i(i)}{\alpha_a(i)} \, i \,(1-i) + \frac{4p(i) \delta}{\alpha_a(i)}i } \right).\notag
\end{eqnarray}
}

Note that for $i = 1 - \frac{\delta}{\beta}$, we get
\begin{equation}\label{eqn:i=1-delta/beta}
a_1\left(1 - \frac{\delta}{\beta}\right) = 0 < a_2\left(1 - \frac{\delta}{\beta}\right).
\end{equation}

All biologically feasible interior equilibria must have their coordinate~$i^*$ in the interval~$\left(0, 1 - \frac{\delta}{\beta}\right)$, as the graph of the linear function~$a_1(i)$ cannot intersect~$int(\Omega)$ for other values of~$i$.

Note also that in this terminology Equation~\eqref{eqn:P3-cond} is equivalent to
\begin{equation}\label{eqn:P3-cond-alt}
a_1(0) > a_2 (0).
\end{equation}

Under the assumptions of part~(a), this guarantees the existence of at least one $i^* \in \left(0, 1 - \frac{\delta}{\beta}\right)$ with $a_1(i^*) = a_2(i^*)$.
Moreover, the intersection of the graphs of~$a_1(i), a_2(i)$ occurs in~$int(\Omega)$. For $\delta \geq \beta_a$ this immediately follows because $a_1(i)$ lies entirely inside $\Omega$ for $i \in  (0,1-\delta/\beta)$ since $a_1(0) < 1$.

For $\delta < \beta_a$, notice that $a_2(1) \geq 0$ and consider the values~$i_0, i_1$ such that $a_2(i)$ enters $\Omega$ crossing the boundary $a+i=1$
at $(i_0,1-i_0)$, and $(i_1,1-i_1)$ is the intersection point of $a_1(i)$ with the same boundary. After substituting $a(i_0) = 1-i_0$ into the left-hand side
of~\eqref{eqn:a-null-implicit}, solving for~$i_0$, and simplifying we obtain:
$$
i_0=\frac{\beta_a - \delta_a -p \delta + \sqrt{(\beta_a - \delta_a -p \delta )^2+4\beta_a\delta_a}}{2\beta_a} > i_1=1-\frac{\delta}{\beta_a}.
$$
Therefore, the graphs of~$a_1(i)$ and~$a_2(i)$ must intersect on the part of the graph of~$a_1(i)$ that connects $\left(1 - \frac{\delta}{\beta}, 0\right)$ and
$(i_1,1-i_1)$, inside of~$\Omega$.

To show uniqueness, under the assumptions of~(a1), consider the function obtained by substituting the above expression for~$a_1(i)$ into the left-hand side of~\eqref{eqn:a-null-implicit}.
This substitution results in a
quadratic function $Q(i)$ that can have at most two roots. The combination of~\eqref{eqn:i=1-delta/beta} and~\eqref{eqn:P3-cond-alt} immediately
rules out the existence of $0 < i^* < i^{**} < 1 - \frac{\delta}{\beta}$ at which the graphs of the functions $a_1(i)$ and~$a_2(i)$  would actually cross.
Moreover, if the the graphs were tangential at~$i^\circ \in \{i^*, i^{**}\}$, then~$i^\circ$ would be a root of~$Q$ of multiplicity~2, and the other equilibrium would need to be a third root of~$Q$, which is impossible.

For the proof of part (b1), consider the limiting case where $\alpha_i = p = 0$.  While we have explicitly ruled it out in our model assumptions, it will be useful in our argument.  In this case, after substituting the expression for $a_1(i)$ in~\eqref{eqn:i-nullcline-inv} into the left hand side of~\eqref{eqn:a-null-implicit}, we obtain a quadratic function in $i$:
\begin{equation*}
    H(i) = \left(\frac{\beta-\delta}{\beta - \beta_a} - \frac{\beta}{\beta-\beta_a}i \right)^2 - \left(1 - i - \frac{\beta_a i + \delta_a}{\alpha_a} \right)\left(\frac{\beta-\delta}{\beta-\beta_a} - \frac{\beta}{\beta - \beta_a}i \right).
\end{equation*}
Then there exists an interior equilibrium if and only if $H(i) = 0$ for some $0 < i < 1 - \frac{\delta}{\beta}$.
By directly evaluating $H(i)$ at $i = 0$ and at $i = 1 - \frac{\delta}{\beta}$, we get
\begin{equation}\label{eqn:H-endpoints}
    H(0) \leq 0\ \ \ \mbox{and}\ \ \ H\left(1-\frac{\delta}{\beta}\right) = 0,
\end{equation}
where the first part relies on the assumption that inequality in~\eqref{eqn:P3-cond} is reversed.

By evaluating $H'(i)$ at $i = 1 - \frac{\delta}{\beta}$, we get
\begin{equation*}
    H'\left(1-\frac{\delta}{\beta}\right) = \frac{\alpha_a\delta + \beta_a\delta - \beta\beta_a - \beta\delta_a}{\alpha_a(\beta-\beta_a)}.
\end{equation*}
Under the assumptions of our model, the denominator is positive,  and inequality~\eqref{eqn:b1-cond} is equivalent to $H'\left(1-\frac{\delta}{\beta}\right) \geq 0$.  It then follows from~\eqref{eqn:H-endpoints} that $H(i) < 0$ for all $0 < i < 1-\frac{\delta}{\beta}$.  Hence, the graphs of $a_1(i)$ and $a_2(i)$ do not intersect in $int(\Omega)$.  Since the inequality~\eqref{eqn:P3-cond-alt} is assumed to be reversed,
we can infer
\begin{equation}\label{eqn:a2i>a1i-int}
    a_2(i) > a_1(i)\ \ \ \mbox{for all}\ \ 0 < i < 1 - \frac{\delta}{\beta}.
\end{equation}

When $\alpha_i + p > 0$, compared to the limiting case above where $\alpha_i = p = 0$, the expression for $a_1(i)$ stays the same while $a_2(i)$ becomes larger for each $0 < i < 1 - \frac{\delta}{\beta}$.  That is, in $int(\Omega)$,  when $\alpha_i + p > 0$, we still have~\eqref{eqn:a2i>a1i-int}.

Therefore, there is no interior equilibrium.

\medskip

For the proof of (a2), suppose $(i^*, a^*)$ is an interior equilibrium, which must be a point on the line segment $a_1(i)$ for some $0 < i^* < 1 - \frac{\delta}{\beta}$.
On this line segment, $a(i)$ is decreasing, and~$s(i) := 1 - a(i) - i$ is increasing, as can be seen if we rewrite~\eqref{eqn:i-nullcline-inv} in its equivalent form given
in~\eqref{eqn:i-null-(a)} \ \ $i(a) = 1 - \frac{\delta}{\beta} - \left(1-\frac{\beta_a}{\beta}\right)a$:
\begin{equation}\label{expr:a_s}
    s(i) =  1 - a(i) - \left[1 - \frac{\delta}{\beta} - \left(1-\frac{\beta_a}{\beta}\right)a(i) \right] = \frac{\delta}{\beta} - \frac{\beta_a}{\beta}a(i).
\end{equation}

We dropped the subscript here since we are only considering $a(i) = a_1(i)$ in this proof.  Now we can rewrite the right hand side of the first line
of~\eqref{rSAISa},  treating $a$ and $s$ as functions of $i$ as in~\eqref{eqn:i-nullcline-inv} and~\eqref{expr:a_s}, to define the following function:
\begin{equation*}
    \begin{split}
        F(i) &= \alpha_i(i)s(i)i + \alpha_a(i)s(i)a(i) + p(i)\delta i - \beta_a a(i) i - \delta_a(i)a(i)\\
        &= \left(\alpha_i(i)\frac{s(i)i}{a(i)} + \alpha_a(i)s(i) - \beta_a i - \delta_a(i) + \frac{p(i)\delta i}{a(i)}\right)a(i)\\
        &= \left[\alpha_i(i)\frac{\frac{\delta}{\beta}-\frac{\beta_a}{\beta}a(i)}{a(i)}i + \alpha_a(i)\left(\frac{\delta}{\beta} - \frac{\beta_a}{\beta}a(i)\right) - \beta_a i - \delta_a(i) + \frac{p(i)\delta i}{a(i)}\right]a(i) .
    \end{split}
\end{equation*}

Let~$f(i)$ be the first factor so that $F(i) = f(i)a(i)$.  It can be expressed as:
\begin{equation}\label{a2:f}
    \begin{split}
        f(i) &= \alpha_i(i)g(i)i + \alpha_a(i)h(i) - \delta_a(i) + \frac{p(i)\delta i}{a(i)},\ \mbox{where}\\
        g(i) &= \frac{\frac{\delta}{\beta} - \frac{\beta_a}{\beta}a(i)}{a(i)} = \frac{\delta}{\beta a(i)} - \frac{\beta_a}{\beta},\\
        h(i) &= \frac{\delta}{\beta} - \frac{\beta_a}{\beta}a(i) - \frac{\beta_a}{\alpha_a(i)}i\\
        &= \frac{\delta}{\beta} - \beta_a\left(\frac{a(i)}{\beta} + \frac{i}{\alpha_a(i)}\right)\\
        &= \frac{\delta}{\beta} - \beta_a\left(\frac{\frac{\beta-\delta}{\beta-\beta_a}-\frac{\beta}{\beta-\beta_a}i}{\beta} + \frac{i}{\alpha_a(i)}\right)\\
        &= \frac{\delta}{\beta} - \frac{\beta_a}{\beta}\left(\frac{\beta-\delta}{\beta-\beta_a}-\frac{\beta}{\beta-\beta_a}i + \frac{\beta}{\alpha_a(i)}i\right)\\
        &= \frac{\delta - \beta_a}{\beta - \beta_a} + \beta_a i\left(\frac{1}{\beta-\beta_a} - \frac{1}{\alpha_a(i)}\right).
    \end{split}
\end{equation}
Then at $(i^*, a^*)$ we must have $f(i^*) = 0$.   Under the assumptions of (a2), $g(i)$ and $h(i)$ are increasing functions in $i$ that take positive values
at any interior equilibrium.
 Hence $f(i)$ is a strictly increasing function in $i$ and we can rule out the existence of two interior equilibria.

\medskip

The result in part~(b2) follows now by the same argument that was used in the proof of part~(a1):  Under the assumptions of (b2),
\begin{equation*}
    \begin{split}
        a_1\left(1-\frac{\delta}{\beta}\right) &< a_2\left(1 - \frac{\delta}{\beta}\right),\\
        a_1(0) &< a_2(0).
    \end{split}
\end{equation*}

Then an interior equilibrium could exist only if the graphs of $a_1(i)$ and $a_2(i)$ would cross at two or more of them, or if $a_1(i)$ and $a_2(i)$ are tangential at one such equilibrium.   The former is ruled out by our uniqueness argument for part~(a2). Note that this argument does not rely on inequality~\eqref{eqn:P3-cond}, which is only needed to prove existence.
Situations where the graphs would cross at one equilibrium~$i^*$ and be tangential at another equilibrium~$i^{**}$  can also be ruled out on purely geometrical grounds: Note that~$a_2(i)$ does not depend on~$\beta$, but~$a_1(i)$ does.  When we slightly increase  $\beta$, both~$a_1(0)$ and $1 - \frac{\delta}{\beta}$  increase. Thus by slightly altering~$\beta$ without violating the assumptions of~(a2), we would obtain a parameter setting with  three interior equilibria, which we have already shown to be impossible.

\medskip

For the proof of~(a3), consider again the function $f(i)$ defined in \eqref{a2:f} and rearrange its terms in the following way:
\begin{equation*}
    \begin{split}
        f(i) &= \alpha_i(i)\frac{\frac{\delta}{\beta} - \frac{\beta_a}{\beta}a(i)}{a(i)}i + \alpha_a(i)\left(\frac{\delta}{\beta} - \frac{\beta_a}{\beta}a(i)\right) - \beta_a i - \delta_a(i) + \frac{p(i)\delta i}{a(i)}\\
        &= \frac{\delta - \beta_a a(i)}{\beta}\left(\frac{\alpha_i(i)}{a(i)}i + \alpha_a(i)\right) + \frac{p(i)\delta i}{a(i)} - \beta_a i - \delta_a(i).
    \end{split}
\end{equation*}
Note that under the assumptions of (a3), the sum of the terms that enter this expression with a positive sign is strictly increasing in $i$, and the sum of the terms $-\beta_a i - \delta_a(i)$ is nondecreasing.  Thus, $f(i)$ is strictly increasing in $i$ and we can rule out the existence of two interior equilibria.

\medskip

The derivation of part~(b3) from part~(a3) is exactly analogous to the derivation of part~(b2) from part~(a2).
$\Box$

\bigskip

\noindent
\textit{Remark}.
Note that the argument in the proof of (b1) relies only on the sign of~$H'\left(1-\frac{\delta}{\beta}\right)$.   Additional sufficient conditions that preclude the existence of interior equilibria could be derived by considering other properties of $H(i)$ or of its counterpart~$G(i)$ that allows for $\alpha_i + p > 0$. Our lemma is not meant to be exhaustive in this respect.
For example, the leading coefficient of the quadratic function~$H(i)$ can be written as
\begin{equation*}
\frac{\beta}{\beta - \beta_a}\left(\frac{\beta}{\beta - \beta_a} - 1 - \frac{\beta_a}{\alpha_a}\right),
\end{equation*}
and is positive if, and only if, $\alpha_a > \beta - \beta_a$, which together with~\eqref{eqn:H-endpoints} gives a proof of~(b2) for the special case of a constant function~$\alpha_a(i)$.

\end{document}